%
%
%

\ifx\mnmacrosloaded\undefined 
%
%
%
%

\catcode `\@=11 

\def\@version{1.6}
\def\@verdate{18th September 1995}

%
%


\newif\ifprod@font

\ifx\@typeface\undefined
  \def\@typeface{Comp. Modern}\prod@fontfalse
\else
  \prod@fonttrue 
\fi

\def\newfam{\alloc@8\fam\chardef\sixt@@n} 

\ifprod@font
\font\fiverm=mtr10 at 5pt
\font\fivebf=mtbx10 at 5pt
\font\fiveit=mtti10 at 5pt
\font\fivesl=mtsl10 at 5pt
\font\fivett=cmtt8 at 5pt     \hyphenchar\fivett=-1
\font\fivecsc=mtcsc10 at 5pt
\font\fivesf=mtss10 at 5pt
\font\fivei=mtmi10 at 5pt      \skewchar\fivei='177
\font\fivesy=mtsy10 at 5pt     \skewchar\fivesy='60

\font\sixrm=mtr10 at 6pt
\font\sixbf=mtbx10 at 6pt
\font\sixit=mtti10 at 6pt
\font\sixsl=mtsl10 at 6pt
\font\sixtt=cmtt8 at 6pt      \hyphenchar\sixtt=-1
\font\sixcsc=mtcsc10 at 6pt
\font\sixsf=mtss10 at 6pt
\font\sixi=mtmi10 at 6pt       \skewchar\sixi='177
\font\sixsy=mtsy10 at 6pt      \skewchar\sixsy='60

\font\sevenrm=mtr10 at 7pt
\font\sevenbf=mtbx10 at 7pt
\font\sevenit=mtti10 at 7pt
\font\sevensl=mtsl10 at 7pt
\font\seventt=cmtt8 at 7pt     \hyphenchar\seventt=-1
\font\sevencsc=mtcsc10 at 7pt
\font\sevensf=mtss10 at 7pt
\font\seveni=mtmi10 at 7pt      \skewchar\seveni='177
\font\sevensy=mtsy10 at 7pt     \skewchar\sevensy='60

\font\eightrm=mtr10 at 8pt
\font\eightbf=mtbx10 at 8pt
\font\eightit=mtti10 at 8pt
\font\eighti=mtmi10 at 8pt      \skewchar\eighti='177
\font\eightsy=mtsy10 at 8pt     \skewchar\eightsy='60
\font\eightsl=mtsl10 at 8pt
\font\eighttt=cmtt8             \hyphenchar\eighttt=-1
\font\eightcsc=mtcsc10 at 8pt
\font\eightsf=mtss10 at 8pt

\font\ninerm=mtr10 at 9pt
\font\ninebf=mtbx10 at 9pt
\font\nineit=mtti10 at 9pt
\font\ninei=mtmi10 at 9pt      \skewchar\ninei='177
\font\ninesy=mtsy10 at 9pt     \skewchar\ninesy='60
\font\ninesl=mtsl10 at 9pt
\font\ninett=cmtt9             \hyphenchar\ninett=-1
\font\ninecsc=mtcsc10 at 9pt
\font\ninesf=mtss10 at 9pt

\font\tenrm=mtr10
\font\tenbf=mtbx10
\font\tenit=mtti10
\font\teni=mtmi10		\skewchar\teni='177
\font\tensy=mtsy10		\skewchar\tensy='60
\font\tenex=cmex10
\font\tensl=mtsl10
\font\tentt=cmtt10		\hyphenchar\tentt=-1
\font\tencsc=mtcsc10
\font\tensf=mtss10

\font\elevenrm=mtr10 at 11pt
\font\elevenbf=mtbx10 at 11pt
\font\elevenit=mtti10 at 11pt
\font\eleveni=mtmi10 at 11pt      \skewchar\eleveni='177
\font\elevensy=mtsy10 at 11pt     \skewchar\elevensy='60
\font\elevensl=mtsl10 at 11pt
\font\eleventt=cmtt10 at 11pt     \hyphenchar\eleventt=-1
\font\elevencsc=mtcsc10 at 11pt
\font\elevensf=mtss10 at 11pt

\font\twelverm=mtr10 at 12pt
\font\twelvebf=mtbx10 at 12pt
\font\twelveit=mtti10 at 12pt
\font\twelvesl=mtsl10 at 12pt
\font\twelvett=cmtt12             \hyphenchar\twelvett=-1
\font\twelvecsc=mtcsc10 at 12pt
\font\twelvesf=mtss10 at 12pt
\font\twelvei=mtmi10 at 12pt      \skewchar\twelvei='177
\font\twelvesy=mtsy10 at 12pt     \skewchar\twelvesy='60

\font\fourteenrm=mtr10 at 14pt
\font\fourteenbf=mtbx10 at 14pt
\font\fourteenit=mtti10 at 14pt
\font\fourteeni=mtmi10 at 14pt      \skewchar\fourteeni='177
\font\fourteensy=mtsy10 at 14pt     \skewchar\fourteensy='60
\font\fourteensl=mtsl10 at 14pt
\font\fourteentt=cmtt12 at 14pt     \hyphenchar\fourteentt=-1
\font\fourteencsc=mtcsc10 at 14pt
\font\fourteensf=mtss10 at 14pt

\font\seventeenrm=mtr10 at 17pt
\font\seventeenbf=mtbx10 at 17pt
\font\seventeenit=mtti10 at 17pt
\font\seventeeni=mtmi10 at 17pt      \skewchar\seventeeni='177
\font\seventeensy=mtsy10 at 17pt     \skewchar\seventeensy='60
\font\seventeensl=mtsl10 at 17pt
\font\seventeentt=cmtt12 at 17pt     \hyphenchar\seventeentt=-1
\font\seventeencsc=mtcsc10 at 17pt
\font\seventeensf=mtss10 at 17pt
\else
\font\fiverm=cmr5
\font\fivei=cmmi5             \skewchar\fivei='177
\font\fivesy=cmsy5            \skewchar\fivesy='60
\font\fivebf=cmbx5

\font\sixrm=cmr6
\font\sixi=cmmi6             \skewchar\sixi='177
\font\sixsy=cmsy6            \skewchar\sixsy='60
\font\sixbf=cmbx6

\font\sevenrm=cmr7
\font\sevenit=cmti7
\font\seveni=cmmi7             \skewchar\seveni='177
\font\sevensy=cmsy7            \skewchar\sevensy='60
\font\sevenbf=cmbx7

\font\eightrm=cmr8
\font\eightbf=cmbx8
\font\eightit=cmti8
\font\eighti=cmmi8			\skewchar\eighti='177
\font\eightsy=cmsy8			\skewchar\eightsy='60
\font\eightsl=cmsl8
\font\eighttt=cmtt8			\hyphenchar\eighttt=-1
\font\eightcsc=cmcsc10 at 8pt
\font\eightsf=cmss8

\font\ninerm=cmr9
\font\ninebf=cmbx9
\font\nineit=cmti9
\font\ninei=cmmi9			\skewchar\ninei='177
\font\ninesy=cmsy9			\skewchar\ninesy='60
\font\ninesl=cmsl9
\font\ninett=cmtt9			\hyphenchar\ninett=-1
\font\ninecsc=cmcsc10 at 9pt
\font\ninesf=cmss9

\font\tenrm=cmr10
\font\tenbf=cmbx10
\font\tenit=cmti10
\font\teni=cmmi10		\skewchar\teni='177
\font\tensy=cmsy10		\skewchar\tensy='60
\font\tenex=cmex10
\font\tensl=cmsl10
\font\tentt=cmtt10		\hyphenchar\tentt=-1
\font\tencsc=cmcsc10
\font\tensf=cmss10

\font\elevenrm=cmr10 scaled \magstephalf
\font\elevenbf=cmbx10 scaled \magstephalf
\font\elevenit=cmti10 scaled \magstephalf
\font\eleveni=cmmi10 scaled \magstephalf	\skewchar\eleveni='177
\font\elevensy=cmsy10 scaled \magstephalf	\skewchar\elevensy='60
\font\elevensl=cmsl10 scaled \magstephalf
\font\eleventt=cmtt10 scaled \magstephalf	\hyphenchar\eleventt=-1
\font\elevencsc=cmcsc10 scaled \magstephalf
\font\elevensf=cmss10 scaled \magstephalf

\font\twelverm=cmr10 scaled \magstep1
\font\twelvebf=cmbx10 scaled \magstep1
\font\twelvei=cmmi10 scaled \magstep1      \skewchar\twelvei='177
\font\twelvesy=cmsy10 scaled \magstep1     \skewchar\twelvesy='60

\font\fourteenrm=cmr10 scaled \magstep2
\font\fourteenbf=cmbx10 scaled \magstep2
\font\fourteenit=cmti10 scaled \magstep2
\font\fourteeni=cmmi10 scaled \magstep2		\skewchar\fourteeni='177
\font\fourteensy=cmsy10 scaled \magstep2	\skewchar\fourteensy='60
\font\fourteensl=cmsl10 scaled \magstep2
\font\fourteentt=cmtt10 scaled \magstep2	\hyphenchar\fourteentt=-1
\font\fourteencsc=cmcsc10 scaled \magstep2
\font\fourteensf=cmss10 scaled \magstep2

\font\seventeenrm=cmr10 scaled \magstep3
\font\seventeenbf=cmbx10 scaled \magstep3
\font\seventeenit=cmti10 scaled \magstep3
\font\seventeeni=cmmi10 scaled \magstep3	\skewchar\seventeeni='177
\font\seventeensy=cmsy10 scaled \magstep3	\skewchar\seventeensy='60
\font\seventeensl=cmsl10 scaled \magstep3
\font\seventeentt=cmtt10 scaled \magstep3	\hyphenchar\seventeentt=-1
\font\seventeencsc=cmcsc10 scaled \magstep3
\font\seventeensf=cmss10 scaled \magstep3
\fi

\def\hexnumber#1{\ifcase#1 0\or1\or2\or3\or4\or5\or6\or7\or8\or9\or
  A\or B\or C\or D\or E\or F\fi}

\def\makestrut{%
  \setbox\strutbox=\hbox{%
    \vrule height.7\baselineskip depth.3\baselineskip width \z@}%
}

\def\baselinestretch{1}
\newskip\tmp@bls

\def\b@ls#1{
  \tmp@bls=#1\relax
  \baselineskip=#1\relax\makestrut
  \normalbaselineskip=\baselinestretch\tmp@bls
  \normalbaselines
}

\def\nostb@ls#1{
  \normalbaselineskip=#1\relax
  \normalbaselines
  \makestrut
}

%

\newfam\scfam  
\newfam\sffam  

\def\mit{\fam\@ne}
\def\cal{\fam\tw@}
\def\em{\ifdim\fontdimen1\font>\z@ \rm\else\it\fi}

\textfont3=\tenex
\scriptfont3=\tenex
\scriptscriptfont3=\tenex

\setbox0=\hbox{\tenex B} 

\def\eightpoint{
  \def\rm{\fam0\eightrm}%
  \textfont0=\eightrm \scriptfont0=\sixrm \scriptscriptfont0=\fiverm%
  \textfont1=\eighti  \scriptfont1=\sixi  \scriptscriptfont1=\fivei%
  \textfont2=\eightsy \scriptfont2=\sixsy \scriptscriptfont2=\fivesy%
  \textfont\itfam=\eightit\def\it{\fam\itfam\eightit}%
  \ifprod@font
    \scriptfont\itfam=\sixit
      \scriptscriptfont\itfam=\fiveit
  \else
    \scriptfont\itfam=\eightit
      \scriptscriptfont\itfam=\eightit
  \fi
  \textfont\bffam=\eightbf%
    \scriptfont\bffam=\sixbf%
      \scriptscriptfont\bffam=\fivebf%
  \def\bf{\fam\bffam\eightbf}%
  \textfont\slfam=\eightsl\def\sl{\fam\slfam\eightsl}%
  \ifprod@font
    \scriptfont\slfam=\sixsl
      \scriptscriptfont\slfam=\fivesl
  \else
    \scriptfont\slfam=\eightsl
      \scriptscriptfont\slfam=\eightsl
  \fi
  \textfont\ttfam=\eighttt\def\tt{\fam\ttfam\eighttt}%
  \ifprod@font
    \scriptfont\ttfam=\sixtt
      \scriptscriptfont\ttfam=\fivett
  \else
    \scriptfont\ttfam=\eighttt
      \scriptscriptfont\ttfam=\eighttt
  \fi
  \textfont\scfam=\eightcsc\def\sc{\fam\scfam\eightcsc}%
  \ifprod@font
    \scriptfont\scfam=\sixcsc
      \scriptscriptfont\scfam=\fivecsc
  \else
    \scriptfont\scfam=\eightcsc
      \scriptscriptfont\scfam=\eightcsc
  \fi
  \textfont\sffam=\eightsf\def\sf{\fam\sffam\eightsf}%
  \ifprod@font
    \scriptfont\sffam=\sixsf
      \scriptscriptfont\sffam=\fivesf
  \else
    \scriptfont\sffam=\eightsf
      \scriptscriptfont\sffam=\eightsf
  \fi
  \def\oldstyle{\fam\@ne\eighti}%
  \b@ls{10pt}\rm\@viiipt%
}
\def\@viiipt{}

\def\ninepoint{
  \def\rm{\fam0\ninerm}%
  \textfont0=\ninerm \scriptfont0=\sixrm \scriptscriptfont0=\fiverm%
  \textfont1=\ninei  \scriptfont1=\sixi  \scriptscriptfont1=\fivei%
  \textfont2=\ninesy \scriptfont2=\sixsy \scriptscriptfont2=\fivesy%
  \textfont\itfam=\nineit\def\it{\fam\itfam\nineit}%
  \ifprod@font
    \scriptfont\itfam=\sixit
      \scriptscriptfont\itfam=\fiveit
  \else
    \scriptfont\itfam=\nineit
      \scriptscriptfont\itfam=\nineit
  \fi
  \textfont\bffam=\ninebf%
    \scriptfont\bffam=\sixbf%
      \scriptscriptfont\bffam=\fivebf%
  \def\bf{\fam\bffam\ninebf}%
  \textfont\slfam=\ninesl\def\sl{\fam\slfam\ninesl}%
  \ifprod@font
    \scriptfont\slfam=\sixsl
      \scriptscriptfont\slfam=\fivesl
  \else
    \scriptfont\slfam=\ninesl
      \scriptscriptfont\slfam=\ninesl
  \fi
  \textfont\ttfam=\ninett\def\tt{\fam\ttfam\ninett}%
  \ifprod@font
    \scriptfont\ttfam=\sixtt
      \scriptscriptfont\ttfam=\fivett
  \else
    \scriptfont\ttfam=\ninett
      \scriptscriptfont\ttfam=\ninett
  \fi
  \textfont\scfam=\ninecsc\def\sc{\fam\scfam\ninecsc}%
  \ifprod@font
    \scriptfont\scfam=\sixcsc
      \scriptscriptfont\scfam=\fivecsc
  \else
    \scriptfont\scfam=\ninecsc
      \scriptscriptfont\scfam=\ninecsc
  \fi
  \textfont\sffam=\ninesf\def\sf{\fam\sffam\ninesf}%
  \ifprod@font
    \scriptfont\sffam=\sixsf
      \scriptscriptfont\sffam=\fivesf
  \else
    \scriptfont\sffam=\ninesf
      \scriptscriptfont\sffam=\ninesf
  \fi
  \def\oldstyle{\fam\@ne\ninei}%
  \b@ls{\TextLeading plus \Feathering}\rm\@ixpt%
}
\def\@ixpt{}

\def\tenpoint{
  \def\rm{\fam0\tenrm}%
  \textfont0=\tenrm \scriptfont0=\sevenrm \scriptscriptfont0=\fiverm%
  \textfont1=\teni  \scriptfont1=\seveni  \scriptscriptfont1=\fivei%
  \textfont2=\tensy \scriptfont2=\sevensy \scriptscriptfont2=\fivesy%
  \textfont\itfam=\tenit\def\it{\fam\itfam\tenit}%
  \ifprod@font
    \scriptfont\itfam=\sevenit
      \scriptscriptfont\itfam=\fiveit
  \else
    \scriptfont\itfam=\tenit
      \scriptscriptfont\itfam=\tenit
  \fi
  \textfont\bffam=\tenbf%
    \scriptfont\bffam=\sevenbf%
      \scriptscriptfont\bffam=\fivebf%
  \def\bf{\fam\bffam\tenbf}%
  \textfont\slfam=\tensl\def\sl{\fam\slfam\tensl}%
  \ifprod@font
    \scriptfont\slfam=\sevensl
      \scriptscriptfont\slfam=\fivesl
  \else
    \scriptfont\slfam=\tensl
      \scriptscriptfont\slfam=\tensl
  \fi
  \textfont\ttfam=\tentt\def\tt{\fam\ttfam\tentt}%
  \ifprod@font
    \scriptfont\ttfam=\seventt
      \scriptscriptfont\ttfam=\fivett
  \else
    \scriptfont\ttfam=\tentt
      \scriptscriptfont\ttfam=\tentt
  \fi
  \textfont\scfam=\tencsc\def\sc{\fam\scfam\tencsc}%
  \ifprod@font
    \scriptfont\scfam=\sevencsc
      \scriptscriptfont\scfam=\fivecsc
  \else
    \scriptfont\scfam=\tencsc
      \scriptscriptfont\scfam=\tencsc
  \fi
  \textfont\sffam=\tensf\def\sf{\fam\sffam\tensf}%
  \ifprod@font
    \scriptfont\sffam=\sevensf
      \scriptscriptfont\sffam=\fivesf
  \else
    \scriptfont\sffam=\tensf
      \scriptscriptfont\sffam=\tensf
  \fi
  \def\oldstyle{\fam\@ne\teni}%
  \b@ls{11pt}\rm\@xpt%
}
\def\@xpt{}

\def\elevenpoint{
  \def\rm{\fam0\elevenrm}%
  \textfont0=\elevenrm \scriptfont0=\eightrm \scriptscriptfont0=\sixrm%
  \textfont1=\eleveni  \scriptfont1=\eighti  \scriptscriptfont1=\sixi%
  \textfont2=\elevensy \scriptfont2=\eightsy \scriptscriptfont2=\sixsy%
  \textfont\itfam=\elevenit\def\it{\fam\itfam\elevenit}%
  \ifprod@font
    \scriptfont\itfam=\eightit
      \scriptscriptfont\itfam=\sixit
  \else
    \scriptfont\itfam=\elevenit
      \scriptscriptfont\itfam=\elevenit
  \fi
  \textfont\bffam=\elevenbf%
    \scriptfont\bffam=\eightbf%
      \scriptscriptfont\bffam=\sixbf%
  \def\bf{\fam\bffam\elevenbf}%
  \textfont\slfam=\elevensl\def\sl{\fam\slfam\elevensl}%
  \ifprod@font
    \scriptfont\slfam=\eightsl
      \scriptscriptfont\slfam=\sixsl
  \else
    \scriptfont\slfam=\elevensl
      \scriptscriptfont\slfam=\elevensl
  \fi
  \textfont\ttfam=\eleventt\def\tt{\fam\ttfam\eleventt}%
  \ifprod@font
    \scriptfont\ttfam=\eighttt
      \scriptscriptfont\ttfam=\sixtt
  \else
    \scriptfont\ttfam=\eleventt
      \scriptscriptfont\ttfam=\eleventt
  \fi
  \textfont\scfam=\elevencsc\def\sc{\fam\scfam\elevencsc}%
  \ifprod@font
    \scriptfont\scfam=\eightcsc
      \scriptscriptfont\scfam=\sixcsc
  \else
    \scriptfont\scfam=\elevencsc
      \scriptscriptfont\scfam=\elevencsc
  \fi
  \textfont\sffam=\elevensf\def\sf{\fam\sffam\elevensf}%
  \ifprod@font
    \scriptfont\sffam=\eightsf
      \scriptscriptfont\sffam=\sixsf
  \else
    \scriptfont\sffam=\elevensf
      \scriptscriptfont\sffam=\elevensf
  \fi
  \def\oldstyle{\fam\@ne\eleveni}%
  \b@ls{13pt}\rm\@xipt%
}
\def\@xipt{}

\def\fourteenpoint{
  \def\rm{\fam0\fourteenrm}%
  \textfont0\fourteenrm  \scriptfont0\tenrm  \scriptscriptfont0\sevenrm%
  \textfont1\fourteeni   \scriptfont1\teni   \scriptscriptfont1\seveni%
  \textfont2\fourteensy  \scriptfont2\tensy  \scriptscriptfont2\sevensy%
  \textfont\itfam=\fourteenit\def\it{\fam\itfam\fourteenit}%
  \ifprod@font
    \scriptfont\itfam=\tenit
      \scriptscriptfont\itfam=\sevenit
  \else
    \scriptfont\itfam=\fourteenit
      \scriptscriptfont\itfam=\fourteenit
  \fi
  \textfont\bffam=\fourteenbf%
    \scriptfont\bffam=\tenbf%
      \scriptscriptfont\bffam=\sevenbf%
  \def\bf{\fam\bffam\fourteenbf}%
  \textfont\slfam=\fourteensl\def\sl{\fam\slfam\fourteensl}%
  \ifprod@font
    \scriptfont\slfam=\tensl
      \scriptscriptfont\slfam=\sevensl
  \else
    \scriptfont\slfam=\fourteensl
      \scriptscriptfont\slfam=\fourteensl
  \fi
  \textfont\ttfam=\fourteentt\def\tt{\fam\ttfam\fourteentt}%
  \ifprod@font
    \scriptfont\ttfam=\tentt
      \scriptscriptfont\ttfam=\seventt
  \else
    \scriptfont\ttfam=\fourteentt
      \scriptscriptfont\ttfam=\fourteentt
  \fi
  \textfont\scfam=\fourteencsc\def\sc{\fam\scfam\fourteencsc}%
  \ifprod@font
    \scriptfont\scfam=\tencsc
      \scriptscriptfont\scfam=\sevencsc
  \else
    \scriptfont\scfam=\fourteencsc
      \scriptscriptfont\scfam=\fourteencsc
  \fi
  \textfont\sffam=\fourteensf\def\sf{\fam\sffam\fourteensf}%
  \ifprod@font
    \scriptfont\sffam=\tensf
      \scriptscriptfont\sffam=\sevensf
  \else
    \scriptfont\sffam=\fourteensf
      \scriptscriptfont\sffam=\fourteensf
  \fi
  \def\oldstyle{\fam\@ne\fourteeni}%
  \b@ls{17pt}\rm\@xivpt%
}
\def\@xivpt{}

\def\seventeenpoint{
  \def\rm{\fam0\seventeenrm}%
  \textfont0\seventeenrm  \scriptfont0\twelverm  \scriptscriptfont0\tenrm%
  \textfont1\seventeeni   \scriptfont1\twelvei   \scriptscriptfont1\teni%
  \textfont2\seventeensy  \scriptfont2\twelvesy  \scriptscriptfont2\tensy%
  \textfont\itfam=\seventeenit\def\it{\fam\itfam\seventeenit}%
  \ifprod@font
    \scriptfont\itfam=\twelveit
      \scriptscriptfont\itfam=\tenit
  \else
    \scriptfont\itfam=\seventeenit
      \scriptscriptfont\itfam=\seventeenit
  \fi
  \textfont\bffam=\seventeenbf%
    \scriptfont\bffam=\twelvebf%
      \scriptscriptfont\bffam=\tenbf%
  \def\bf{\fam\bffam\seventeenbf}%
  \textfont\slfam=\seventeensl\def\sl{\fam\slfam\seventeensl}%
  \ifprod@font
    \scriptfont\slfam=\twelvesl
      \scriptscriptfont\slfam=\tensl
  \else
    \scriptfont\slfam=\seventeensl
      \scriptscriptfont\slfam=\seventeensl
  \fi
  \textfont\ttfam=\seventeentt\def\tt{\fam\ttfam\seventeentt}%
  \ifprod@font
    \scriptfont\ttfam=\twelvett
      \scriptscriptfont\ttfam=\tentt
  \else
    \scriptfont\ttfam=\seventeentt
      \scriptscriptfont\ttfam=\seventeentt
  \fi
  \textfont\scfam=\seventeencsc\def\sc{\fam\scfam\seventeencsc}%
  \ifprod@font
    \scriptfont\scfam=\twelvecsc
      \scriptscriptfont\scfam=\tencsc
  \else
    \scriptfont\scfam=\seventeencsc
      \scriptscriptfont\scfam=\seventeencsc
  \fi
  \textfont\sffam=\seventeensf\def\sf{\fam\sffam\seventeensf}%
  \ifprod@font
    \scriptfont\sffam=\twelvesf
      \scriptscriptfont\sffam=\tensf
  \else
    \scriptfont\sffam=\seventeensf
      \scriptscriptfont\sffam=\seventeensf
  \fi
  \def\oldstyle{\fam\@ne\seventeeni}%
  \b@ls{20pt}\rm\@xviipt%
}
\def\@xviipt{}

\lineskip=1pt      \normallineskip=\lineskip
\lineskiplimit=\z@ \normallineskiplimit=\lineskiplimit


\def\loadboldmathnames{%
  \def\balpha{{\bmath{\alpha}}}%
  \def\bbeta{{\bmath{\beta}}}%
  \def\bgamma{{\bmath{\gamma}}}%
  \def\bdelta{{\bmath{\delta}}}%
  \def\bepsilon{{\bmath{\epsilon}}}%
  \def\bzeta{{\bmath{\zeta}}}%
  \def\boldeta{{\bmath{\eta}}}%
  \def\btheta{{\bmath{\theta}}}%
  \def\biota{{\bmath{\iota}}}%
  \def\bkappa{{\bmath{\kappa}}}%
  \def\blambda{{\bmath{\lambda}}}%
  \def\bmu{{\bmath{\mu}}}%
  \def\bnu{{\bmath{\nu}}}%
  \def\bxi{{\bmath{\xi}}}%
  \def\bpi{{\bmath{\pi}}}%
  \def\brho{{\bmath{\rho}}}%
  \def\bsigma{{\bmath{\sigma}}}%
  \def\btau{{\bmath{\tau}}}%
  \def\bupsilon{{\bmath{\upsilon}}}%
  \def\bphi{{\bmath{\phi}}}%
  \def\bchi{{\bmath{\chi}}}%
  \def\bpsi{{\bmath{\psi}}}%
  \def\bomega{{\bmath{\omega}}}%
  \def\bvarepsilon{{\bmath{\varepsilon}}}%
  \def\bvartheta{{\bmath{\vartheta}}}%
  \def\bvarpi{{\bmath{\varpi}}}%
  \def\bvarrho{{\bmath{\varrho}}}%
  \def\bvarsigma{{\bmath{\varsigma}}}%
  \def\bvarphi{{\bmath{\varphi}}}%
  \def\baleph{{\bmath{\aleph}}}%
  \def\bimath{{\bmath{\imath}}}%
  \def\bjmath{{\bmath{\jmath}}}%
  \def\bell{{\bmath{\ell}}}%
  \def\bwp{{\bmath{\wp}}}%
  \def\bRe{{\bmath{\Re}}}%
  \def\bIm{{\bmath{\Im}}}%
  \def\bpartial{{\bmath{\partial}}}%
  \def\binfty{{\bmath{\infty}}}%
  \def\bprime{{\bmath{\prime}}}%
  \def\bemptyset{{\bmath{\emptyset}}}%
  \def\bnabla{{\bmath{\nabla}}}%
  \def\btop{{\bmath{\top}}}%
  \def\bbot{{\bmath{\bot}}}%
  \def\btriangle{{\bmath{\triangle}}}%
  \def\bforall{{\bmath{\forall}}}%
  \def\bexists{{\bmath{\exists}}}%
  \def\bneg{{\bmath{\neg}}}%
  \def\bflat{{\bmath{\flat}}}%
  \def\bnatural{{\bmath{\natural}}}%
  \def\bsharp{{\bmath{\sharp}}}%
  \def\bclubsuit{{\bmath{\clubsuit}}}%
  \def\bdiamondsuit{{\bmath{\diamondsuit}}}%
  \def\bheartsuit{{\bmath{\heartsuit}}}%
  \def\bspadesuit{{\bmath{\spadesuit}}}%
  \def\bsmallint{{\bmath{\smallint}}}%
  \def\btriangleleft{{\bmath{\triangleleft}}}%
  \def\btriangleright{{\bmath{\triangleright}}}%
  \def\bbigtriangleup{{\bmath{\bigtriangleup}}}%
  \def\bbigtriangledown{{\bmath{\bigtriangledown}}}%
  \def\bwedge{{\bmath{\wedge}}}%
  \def\bvee{{\bmath{\vee}}}%
  \def\bcap{{\bmath{\cap}}}%
  \def\bcup{{\bmath{\cup}}}%
  \def\bddagger{{\bmath{\ddagger}}}%
  \def\bdagger{{\bmath{\dagger}}}%
  \def\bsqcap{{\bmath{\sqcap}}}%
  \def\bsqcup{{\bmath{\sqcup}}}%
  \def\buplus{{\bmath{\uplus}}}%
  \def\bamalg{{\bmath{\amalg}}}%
  \def\bdiamond{{\bmath{\diamond}}}%
  \def\bbullet{{\bmath{\bullet}}}%
  \def\bwr{{\bmath{\wr}}}%
  \def\bdiv{{\bmath{\div}}}%
  \def\bodot{{\bmath{\odot}}}%
  \def\boslash{{\bmath{\oslash}}}%
  \def\botimes{{\bmath{\otimes}}}%
  \def\bominus{{\bmath{\ominus}}}%
  \def\boplus{{\bmath{\oplus}}}%
  \def\bmp{{\bmath{\mp}}}%
  \def\bpm{{\bmath{\pm}}}%
  \def\bcirc{{\bmath{\circ}}}%
  \def\bbigcirc{{\bmath{\bigcirc}}}%
  \def\bsetminus{{\bmath{\setminus}}}%
  \def\bcdot{{\bmath{\cdot}}}%
  \def\bast{{\bmath{\ast}}}%
  \def\btimes{{\bmath{\times}}}%
  \def\bstar{{\bmath{\star}}}%
  \def\bpropto{{\bmath{\propto}}}%
  \def\bsqsubseteq{{\bmath{\sqsubseteq}}}%
  \def\bsqsupseteq{{\bmath{\sqsupseteq}}}%
  \def\bparallel{{\bmath{\parallel}}}%
  \def\bmid{{\bmath{\mid}}}%
  \def\bdashv{{\bmath{\dashv}}}%
  \def\bvdash{{\bmath{\vdash}}}%
  \def\bnearrow{{\bmath{\nearrow}}}%
  \def\bsearrow{{\bmath{\searrow}}}%
  \def\bnwarrow{{\bmath{\nwarrow}}}%
  \def\bswarrow{{\bmath{\swarrow}}}%
  \def\bLeftrightarrow{{\bmath{\Leftrightarrow}}}%
  \def\bLeftarrow{{\bmath{\Leftarrow}}}%
  \def\bRightarrow{{\bmath{\Rightarrow}}}%
  \def\bleq{{\bmath{\leq}}}%
  \def\bgeq{{\bmath{\geq}}}%
  \def\bsucc{{\bmath{\succ}}}%
  \def\bprec{{\bmath{\prec}}}%
  \def\bapprox{{\bmath{\approx}}}%
  \def\bsucceq{{\bmath{\succeq}}}%
  \def\bpreceq{{\bmath{\preceq}}}%
  \def\bsupset{{\bmath{\supset}}}%
  \def\bsubset{{\bmath{\subset}}}%
  \def\bsupseteq{{\bmath{\supseteq}}}%
  \def\bsubseteq{{\bmath{\subseteq}}}%
  \def\bin{{\bmath{\in}}}%
  \def\bni{{\bmath{\ni}}}%
  \def\bgg{{\bmath{\gg}}}%
  \def\bll{{\bmath{\ll}}}%
  \def\bnot{{\bmath{\not}}}%
  \def\bleftrightarrow{{\bmath{\leftrightarrow}}}%
  \def\bleftarrow{{\bmath{\leftarrow}}}%
  \def\brightarrow{{\bmath{\rightarrow}}}%
  \def\bmapstochar{{\bmath{\mapstochar}}}%
  \def\bsim{{\bmath{\sim}}}%
  \def\bsimeq{{\bmath{\simeq}}}%
  \def\bperp{{\bmath{\perp}}}%
  \def\bequiv{{\bmath{\equiv}}}%
  \def\basymp{{\bmath{\asymp}}}%
  \def\bsmile{{\bmath{\smile}}}%
  \def\bfrown{{\bmath{\frown}}}%
  \def\bleftharpoonup{{\bmath{\leftharpoonup}}}%
  \def\bleftharpoondown{{\bmath{\leftharpoondown}}}%
  \def\brightharpoonup{{\bmath{\rightharpoonup}}}%
  \def\brightharpoondown{{\bmath{\rightharpoondown}}}%
  \def\blhook{{\bmath{\lhook}}}%
  \def\brhook{{\bmath{\rhook}}}%
  \def\bldotp{{\bmath{\ldotp}}}%
  \def\bcdotp{{\bmath{\cdotp}}}%
}

\def\,{\relax\ifmmode \mskip\thinmuskip\else \thinspace\fi}
\let\protect=\relax

\long\def\@ifundefined#1#2#3{\expandafter\ifx\csname
  #1\endcsname\relax#2\else#3\fi}




\newtoks\math@groups \math@groups={}
\def\addtom@thgroup#1#2{#1\expandafter{\the#1#2}} 



\def\addtosizeh@ok#1#2#3#4{%
  \expandafter\def\csname @#1pt\endcsname{%
    \def\s@ze{#2}\def\ss@ze{#3}\def\sss@ze{#4}\the\math@groups%
  }%
}



\let\resetsizehook=\addtosizeh@ok


\ifprod@font
  \addtosizeh@ok{viii} {8} {6}  {5}
  \addtosizeh@ok{ix}   {9} {6}  {5}
  \addtosizeh@ok{x}    {10}{7}  {5}
  \addtosizeh@ok{xi}   {11}{8}  {6}
  \addtosizeh@ok{xiv}  {14}{10} {7}
  \addtosizeh@ok{xvii} {17}{12}{10}
\else
  \addtosizeh@ok{viii} {8}     {6}     {5}
  \addtosizeh@ok{ix}   {9}     {6}     {5}
  \addtosizeh@ok{x}    {10}    {7}     {5}
  \addtosizeh@ok{xi}   {10.95} {8}     {6}
  \addtosizeh@ok{xiv}  {14.4}  {10}    {7}
  \addtosizeh@ok{xvii} {17.28} {12}    {10}
\fi

\def\get@font#1#2#3{%
  \edef\fonts@ze{\romannumeral#3}
  \edef\fontn@me{\fonts@ze#1}
  \@ifundefined{\fontn@me}%
    {
     \global\expandafter\font\csname \fontn@me\endcsname=#2 at #3pt}%
    {}%
}

\def\ass@tfont#1#2{%
  \xdef\fam@name{\csname #1\endcsname}%
  \xdef\font@name{\csname #2\endcsname}%
  \let\textfont@name\font@name
  \textfont\fam@name\textfont@name
}

\def\ass@sfont#1#2{%
  \xdef\fam@name{\csname #1\endcsname}%
  \xdef\font@name{\csname #2\endcsname}%
  \let\textfont@name\font@name
  \scriptfont\fam@name\textfont@name
}

\def\ass@ssfont#1#2{%
  \xdef\fam@name{\csname #1\endcsname}%
  \xdef\font@name{\csname #2\endcsname}%
  \let\textfont@name\font@name
  \scriptscriptfont\fam@name\textfont@name
}


\def\NewSymbolFont#1#2{%
  \expandafter\ifx\csname sym#1fam\endcsname\relax 
    \expandafter\newfam\csname sym#1fam\endcsname
    \expandafter\edef\csname sym#1fam\endcsname{\the\allocationnumber}%
    \addtom@thgroup\math@groups{%
      \get@font{#1}{#2}{\s@ze}%
      \ass@tfont{sym#1fam}{\fontn@me}%
      \get@font{#1}{#2}{\ss@ze}%
      \ass@sfont{sym#1fam}{\fontn@me}%
      \get@font{#1}{#2}{\sss@ze}%
      \ass@ssfont{sym#1fam}{\fontn@me}%
    }%
  \else
    \errmessage{Family `#1' already defined}%
  \fi
}


\def\NewMathSymbol#1#2#3#4{%
  \edef\f@mly{\expandafter\hexnumber{\csname sym#3fam\endcsname}}%
  \mathchardef#1="#2\f@mly#4\relax
}


\newif\ifd@f

\def\NewMathDelimiter#1#2#3#4#5#6{%
  \d@ftrue
  \expandafter\ifx\csname sym#3fam\endcsname\relax
    \d@ffalse \errmessage{Family `#3' is not defined}%
  \fi
  \expandafter\ifx\csname sym#5fam\endcsname\relax
    \d@ffalse \errmessage{Family `#5' is not defined}%
  \fi
  \ifd@f
    \edef\f@mly{\expandafter\hexnumber{\csname sym#3fam\endcsname}}%
    \edef\f@mlytw@{\expandafter\hexnumber{\csname sym#5fam\endcsname}}%
    \xdef#1{\delimiter"#2\f@mly #4\f@mlytw@ #6\relax}%
  \fi
}


\def\setboxz@h{\setbox\z@\hbox}
\def\wdz@{\wd\z@}
\def\boxz@{\box\z@}
\def\setbox@ne{\setbox\@ne}
\def\wd@ne{\wd\@ne}

\def\math@atom#1#2{%
   \binrel@{#1}\binrel@@{#2}}
\def\binrel@#1{\setboxz@h{\thinmuskip0mu
  \medmuskip\m@ne mu\thickmuskip\@ne mu$#1\m@th$}%
 \setbox@ne\hbox{\thinmuskip0mu\medmuskip\m@ne mu\thickmuskip
  \@ne mu${}#1{}\m@th$}%
 \setbox\tw@\hbox{\hskip\wd@ne\hskip-\wdz@}}
\def\binrel@@#1{\ifdim\wd2<\z@\mathbin{#1}\else\ifdim\wd\tw@>\z@
 \mathrel{#1}\else{#1}\fi\fi}

\def\m@thit{1}

\def\set@skchar#1{\global\expandafter\skewchar
  \csname\fontn@me\endcsname=#1\relax}

\def\NewMathAlphabet#1#2#3{%
  \def\tst{#3}%
  \ifx\tst\empty\else 
    \expandafter\gdef\csname #1@sc\endcsname{}
  \fi
  \expandafter\def\csname #1\endcsname{
    \protect\csname @#1\endcsname}%
  \expandafter\def\csname @#1\endcsname##1{
    {%
    \begingroup
      \get@font{#1}{#2}{\s@ze}%
      \@ifundefined{#1@sc}{}{\set@skchar{#3}}%
      \ass@tfont{m@thit}{\fontn@me}%
      \get@font{#1}{#2}{\ss@ze}%
      \@ifundefined{#1@sc}{}{\set@skchar{#3}}%
      \ass@sfont{m@thit}{\fontn@me}%
      \get@font{#1}{#2}{\sss@ze}%
      \@ifundefined{#1@sc}{}{\set@skchar{#3}}%
      \ass@ssfont{m@thit}{\fontn@me}%
      \math@atom{##1}{%
      \mathchoice%
        {\hbox{$\m@th\displaystyle##1$}}%
        {\hbox{$\m@th\textstyle##1$}}%
        {\hbox{$\m@th\scriptstyle##1$}}%
        {\hbox{$\m@th\scriptscriptstyle##1$}}}%
    \endgroup
    }%
  }%
}


\newif\iffirstta  \firsttatrue

\def\set@hchar#1{\global\expandafter\hyphenchar
  \csname\fontn@me\endcsname=#1\relax}

\def\NewTextAlphabet#1#2#3{%
  \iffirstta
    \global\firsttafalse
    \newfam\scratchfam
    \edef\scrt@fam{\the\allocationnumber}%
  \fi
  \def\tst{#3}%
  \ifx\tst\empty\else 
    \expandafter\gdef\csname #1@hc\endcsname{}
  \fi
  \expandafter\def\csname #1\endcsname{
    \protect\csname t@#1\endcsname}%
  \long\expandafter\def\csname t@#1\endcsname##1{
    \ifmmode
      \typeout{Warning: do not use \expandafter\string\csname #1\endcsname
        \space in math mode}\fi%
    {%
      \get@font{#1}{#2}{\s@ze}\let\t@xtfnt=\fontn@me\relax
      \@ifundefined{#1@hc}{}{\set@hchar{#3}}%
      \ass@tfont{scrt@fam}{\fontn@me}%
      \get@font{#1}{#2}{\ss@ze}%
      \@ifundefined{#1@hc}{}{\set@hchar{#3}}%
      \ass@sfont{scrt@fam}{\fontn@me}%
      \get@font{#1}{#2}{\sss@ze}%
      \@ifundefined{#1@hc}{}{\set@hchar{#3}}%
      \ass@ssfont{scrt@fam}{\fontn@me}%
      \fam\scratchfam\csname\t@xtfnt\endcsname
    ##1%
    }%
  }%
  \expandafter\def\csname #1shape
    \endcsname{\protect\csname @#1shape\endcsname}%
  \expandafter\def\csname @#1shape\endcsname{
    \ifmmode
      \typeout{Warning: do not use \expandafter\string\csname
        #1shape\endcsname \space in math mode}\fi
      \get@font{#1}{#2}{\s@ze}\let\t@xtfnt=\fontn@me\relax
      \@ifundefined{#1@hc}{}{\set@hchar{#3}}%
      \ass@tfont{scrt@fam}{\fontn@me}%
      \get@font{#1}{#2}{\ss@ze}%
      \@ifundefined{#1@hc}{}{\set@hchar{#3}}%
      \ass@sfont{scrt@fam}{\fontn@me}%
      \get@font{#1}{#2}{\sss@ze}%
      \@ifundefined{#1@hc}{}{\set@hchar{#3}}%
      \ass@ssfont{scrt@fam}{\fontn@me}%
      \fam\scratchfam\csname\t@xtfnt\endcsname
  }%
}


\ifprod@font
  \def\math@itfnt{mtmib10}
  \def\math@syfnt{mtbsy10}
\else
  \def\math@itfnt{cmmib10}
  \def\math@syfnt{cmbsy10}
\fi

\def\m@thsy{2}

\def\bmath{\protect\@bmath}
\def\@bmath#1{%
  {%
  \begingroup
    \get@font{mthit}{\math@itfnt}{\s@ze}\set@skchar{'177}%
    \ass@tfont{m@thit}{\fontn@me}%
    \get@font{mthit}{\math@itfnt}{\ss@ze}\set@skchar{'177}%
    \ass@sfont{m@thit}{\fontn@me}%
    \get@font{mthit}{\math@itfnt}{\sss@ze}\set@skchar{'177}%
    \ass@ssfont{m@thit}{\fontn@me}%
    \get@font{mthsy}{\math@syfnt}{\s@ze}\set@skchar{'60}%
    \ass@tfont{m@thsy}{\fontn@me}%
    \get@font{mthsy}{\math@syfnt}{\ss@ze}\set@skchar{'60}%
    \ass@sfont{m@thsy}{\fontn@me}%
    \get@font{mthsy}{\math@syfnt}{\sss@ze}\set@skchar{'60}%
    \ass@ssfont{m@thsy}{\fontn@me}%
    \math@atom{#1}{%
    \mathchoice%
      {\hbox{$\m@th\displaystyle#1$}}%
      {\hbox{$\m@th\textstyle#1$}}%
      {\hbox{$\m@th\scriptstyle#1$}}%
      {\hbox{$\m@th\scriptscriptstyle#1$}}}%
  \endgroup
  }%
}



\def\diameter{{\ifmmode\mathchoice
{\ooalign{\hfil\hbox{$\displaystyle/$}\hfil\crcr
{\hbox{$\displaystyle\mathchar"20D$}}}}
{\ooalign{\hfil\hbox{$\textstyle/$}\hfil\crcr
{\hbox{$\textstyle\mathchar"20D$}}}}
{\ooalign{\hfil\hbox{$\scriptstyle/$}\hfil\crcr
{\hbox{$\scriptstyle\mathchar"20D$}}}}
{\ooalign{\hfil\hbox{$\scriptscriptstyle/$}\hfil\crcr
{\hbox{$\scriptscriptstyle\mathchar"20D$}}}}
\else{\ooalign{\hfil/\hfil\crcr\mathhexbox20D}}%
\fi}}

\def\sq{\ifmmode\squareforqed\else{\unskip\nobreak\hfil
\penalty50\hskip1em\null\nobreak\hfil\squareforqed
\parfillskip=0pt\finalhyphendemerits=0\endgraf}\fi}
\def\squareforqed{\hbox{\rlap{$\sqcap$}$\sqcup$}}


\def\bbbc{{\mathchoice {\setbox0=\hbox{$\displaystyle\rm C$}\hbox{\hbox
to0pt{\kern0.4\wd0\vrule height0.9\ht0\hss}\box0}}
{\setbox0=\hbox{$\textstyle\rm C$}\hbox{\hbox
to0pt{\kern0.4\wd0\vrule height0.9\ht0\hss}\box0}}
{\setbox0=\hbox{$\scriptstyle\rm C$}\hbox{\hbox
to0pt{\kern0.4\wd0\vrule height0.9\ht0\hss}\box0}}
{\setbox0=\hbox{$\scriptscriptstyle\rm C$}\hbox{\hbox
to0pt{\kern0.4\wd0\vrule height0.9\ht0\hss}\box0}}}}
\def\bbbq{{\mathchoice {\setbox0=\hbox{$\displaystyle\rm
Q$}\hbox{\raise
0.15\ht0\hbox to0pt{\kern0.4\wd0\vrule height0.8\ht0\hss}\box0}}
{\setbox0=\hbox{$\textstyle\rm Q$}\hbox{\raise
0.15\ht0\hbox to0pt{\kern0.4\wd0\vrule height0.8\ht0\hss}\box0}}
{\setbox0=\hbox{$\scriptstyle\rm Q$}\hbox{\raise
0.15\ht0\hbox to0pt{\kern0.4\wd0\vrule height0.7\ht0\hss}\box0}}
{\setbox0=\hbox{$\scriptscriptstyle\rm Q$}\hbox{\raise
0.15\ht0\hbox to0pt{\kern0.4\wd0\vrule height0.7\ht0\hss}\box0}}}}
\def\bbbt{{\mathchoice {\setbox0=\hbox{$\displaystyle\rm
T$}\hbox{\hbox to0pt{\kern0.3\wd0\vrule height0.9\ht0\hss}\box0}}
{\setbox0=\hbox{$\textstyle\rm T$}\hbox{\hbox
to0pt{\kern0.3\wd0\vrule height0.9\ht0\hss}\box0}}
{\setbox0=\hbox{$\scriptstyle\rm T$}\hbox{\hbox
to0pt{\kern0.3\wd0\vrule height0.9\ht0\hss}\box0}}
{\setbox0=\hbox{$\scriptscriptstyle\rm T$}\hbox{\hbox
to0pt{\kern0.3\wd0\vrule height0.9\ht0\hss}\box0}}}}
\def\bbbs{{\mathchoice
{\setbox0=\hbox{$\displaystyle     \rm S$}\hbox{\raise0.5\ht0\hbox
to0pt{\kern0.35\wd0\vrule height0.45\ht0\hss}\hbox
to0pt{\kern0.55\wd0\vrule height0.5\ht0\hss}\box0}}
{\setbox0=\hbox{$\textstyle        \rm S$}\hbox{\raise0.5\ht0\hbox
to0pt{\kern0.35\wd0\vrule height0.45\ht0\hss}\hbox
to0pt{\kern0.55\wd0\vrule height0.5\ht0\hss}\box0}}
{\setbox0=\hbox{$\scriptstyle      \rm S$}\hbox{\raise0.5\ht0\hbox
to0pt{\kern0.35\wd0\vrule height0.45\ht0\hss}\raise0.05\ht0\hbox
to0pt{\kern0.5\wd0\vrule height0.45\ht0\hss}\box0}}
{\setbox0=\hbox{$\scriptscriptstyle\rm S$}\hbox{\raise0.5\ht0\hbox
to0pt{\kern0.4\wd0\vrule height0.45\ht0\hss}\raise0.05\ht0\hbox
to0pt{\kern0.55\wd0\vrule height0.45\ht0\hss}\box0}}}}
\def\bbbz{{\mathchoice {\hbox{$\sf\textstyle Z\kern-0.4em Z$}}
{\hbox{$\sf\textstyle Z\kern-0.4em Z$}}
{\hbox{$\sf\scriptstyle Z\kern-0.3em Z$}}
{\hbox{$\sf\scriptscriptstyle Z\kern-0.2em Z$}}}}


\def\Nulle{0} 
\def\Afe{1}   
\def\Hae{2}   
\def\Hbe{3}   
\def\Hce{4}   
\def\Hde{5}   


\newcount\LastMac       \LastMac=\Nulle

\newskip\half      \half=5.5pt plus 1.5pt minus 2.25pt
\newskip\one       \one=11pt plus 3pt minus 5.5pt
\newskip\onehalf   \onehalf=16.5pt plus 5.5pt minus 8.25pt
\newskip\two       \two=22pt plus 5.5pt minus 11pt

\def\Half{\addvspace{\half}}
\def\One{\addvspace{\one}}
\def\OneHalf{\addvspace{\onehalf}}
\def\Two{\addvspace{\two}}

\def\Raggedright{
  \rightskip=\z@ plus \hsize\relax
}

\def\Fullout{
  \rightskip=\z@\relax
}

\def\Hang#1#2{
  \hangindent=#1%
  \hangafter=#2\relax
}


\newif\ifsp@page
\def\pagestyle#1{\csname ps@#1\endcsname}
\def\thispagestyle#1{\global\sp@pagetrue\gdef\sp@type{#1}}

\def\ps@titlepage{%
  \def\@oddhead{\eightpoint\noindent \the\CatchLine
    \ifprod@font\else\qquad Printed\ \today\qquad
      (MN plain \TeX\ macros\ v\@version)\fi \hfil}%
  \let\@evenhead=\@oddhead
  \def\@oddfoot{\eightpoint\copyright\ \@pubyear\ RAS\hfil}%
  \def\@evenfoot{\hfil\eightpoint\noindent\copyright\ \@pubyear\ RAS}%
}

\def\ps@headings{%
  \def\@oddhead{\elevenpoint\it\noindent
    \hfill\the\RightHeader\hskip1.5em\rm\folio}%
  \def\@evenhead{\elevenpoint\noindent
    \folio\hskip1.5em\it\the\LeftHeader\hfill}%
  \def\@oddfoot{\eightpoint\noindent\copyright\ \@pubyear\ RAS,
    MNRAS {\bf \@volume}, \@pagerange\hfil}%
  \def\@evenfoot{\hfil\eightpoint\copyright\ \@pubyear\ RAS,
    MNRAS {\bf \@volume}, \@pagerange}%
}

\def\ps@plate{%
  \def\@oddhead{\eightpoint\noindent\plt@cap\hfil}%
  \def\@evenhead{\eightpoint\noindent\plt@cap\hfil}%
  \def\@oddfoot{\eightpoint\noindent\copyright\ \@pubyear\ RAS,
    MNRAS {\bf \@volume}, \@pagerange\hfil}%
  \def\@evenfoot{\hfil\eightpoint\copyright\ \@pubyear\ RAS,
    MNRAS {\bf \@volume}, \@pagerange}%
}



\def\title#1{
  \bgroup
    \vbox to 8pt{\vss}%
    \seventeenpoint
    \Raggedright
    \noindent \strut{\bf #1}\par
  \egroup
}

\def\author#1{
  \bgroup
    \ifnum\LastMac=\Afe \OneHalf\else \vskip 21pt\fi
    \fourteenpoint
    \Raggedright
    \noindent \strut #1\par
    \vskip 3pt%
  \egroup
}

\def\affiliation#1{
  \bgroup
    \vskip -4pt%
    \eightpoint
    \Raggedright
    \noindent \strut {\it #1}\par
  \egroup
  \LastMac=\Afe\relax
}

\def\acceptedline#1{
  \bgroup
    \Two
    \eightpoint
    \Raggedright
    \noindent \strut #1\par
  \egroup
}

\long\def\abstract#1{%
  \bgroup
    \vskip 20pt%
    \leftskip 11pc\rightskip\z@
    \noindent{\ninebf ABSTRACT}\par
    \tenpoint
    \Fullout
    \noindent #1\par
  \egroup
}

\long\def\keywords#1{
  \bgroup
    \Half
    \leftskip 11pc\rightskip\z@
    \tenpoint
    \Fullout
    \noindent\hbox{\bf Key words:}\ #1\par
  \egroup
}


\def\maketitle{%
  \EndOpening
  \ifsinglecol \else \MakePage\fi
}


\def\pageoffset#1#2{\hoffset=#1\relax\voffset=#2\relax}


\def\@nameuse#1{\csname #1\endcsname}
\def\arabic#1{\@arabic{\@nameuse{#1}}}
\def\alph#1{\@alph{\@nameuse{#1}}}
\def\Alph#1{\@Alph{\@nameuse{#1}}}
\def\@arabic#1{\number #1}
\def\@Alph#1{\ifcase#1\or A\or B\or C\or D\else\@Ialph{#1}\fi}
\def\@Ialph#1{\ifcase#1\or \or \or \or \or E\or F\or G\or H\or I\or J\or
   K\or L\or M\or N\or O\or P\or Q\or R\or S\or T\or U\or V\or W\or X\or
   Y\or Z\else\errmessage{Counter out of range}\fi}
\def\@alph#1{\ifcase#1\or a\or b\or c\or d\else\@ialph{#1}\fi}
\def\@ialph#1{\ifcase#1\or \or \or \or \or e\or f\or g\or h\or i\or j\or
   k\or l\or m\or n\or o\or p\or q\or r\or s\or t\or u\or v\or w\or x\or y\or
   z\else\errmessage{Counter out of range}\fi}


\newcount\Eqnno
\newcount\SubEqnno

\def\theeq{\arabic{Eqnno}}
\def\thesubeq{\alph{SubEqnno}}

\def\stepeq{\relax
  \global\SubEqnno \z@
  \global\advance\Eqnno \@ne\relax
  {\rm (\theeq)}%
}

\def\startsubeq{\relax
  \global\SubEqnno \z@
  \global\advance\Eqnno \@ne\relax
  \stepsubeq
}

\def\stepsubeq{\relax
  \global\advance\SubEqnno \@ne\relax
  {\rm (\theeq\thesubeq)}%
}


\newcount\Sec        
\newcount\SecSec
\newcount\SecSecSec

\def\thesection{\arabic{Sec}}
\def\thesubsection{\thesection.\arabic{SecSec}}
\def\thesubsubsection{\thesubsection.\arabic{SecSecSec}}

\Sec=\z@

\def\:{\let\@sptoken= } \:  
\def\:{\@xifnch} \expandafter\def\: {\futurelet\@tempc\@ifnch}

\def\@ifnextchar#1#2#3{%
  \let\@tempMACe #1%
  \def\@tempMACa{#2}%
  \def\@tempMACb{#3}%
  \futurelet \@tempMACc\@ifnch%
}

\def\@ifnch{%
\ifx \@tempMACc \@sptoken%
  \let\@tempMACd\@xifnch%
\else%
  \ifx \@tempMACc \@tempMACe%
    \let\@tempMACd\@tempMACa%
  \else%
    \let\@tempMACd\@tempMACb%
  \fi%
\fi%
\@tempMACd%
}

\def\@ifstar#1#2{\@ifnextchar *{\def\@tempMACa*{#1}\@tempMACa}{#2}}

\newskip\@tempskipb

\def\addvspace#1{%
  \ifvmode\else \endgraf\fi%
  \ifdim\lastskip=\z@%
    \vskip #1\relax%
  \else%
    \@tempskipb#1\relax\@xaddvskip%
  \fi%
}

\def\@xaddvskip{%
  \ifdim\lastskip<\@tempskipb%
    \vskip-\lastskip%
    \vskip\@tempskipb\relax%
  \else%
    \ifdim\@tempskipb<\z@%
      \ifdim\lastskip<\z@ \else%
        \advance\@tempskipb\lastskip%
        \vskip-\lastskip\vskip\@tempskipb%
      \fi%
    \fi%
  \fi%
}

\newskip\@tmpSKIP

\def\addpen#1{%
  \ifvmode
    \if@nobreak
    \else
      \ifdim\lastskip=\z@
        \penalty#1\relax
      \else
        \@tmpSKIP=\lastskip
        \vskip -\lastskip
        \penalty#1\vskip\@tmpSKIP
      \fi
    \fi
  \fi
}

\newcount\@clubpen   \@clubpen=\clubpenalty
\newif\if@nobreak    \@nobreakfalse

\def\@noafterindent{%
  \global\@nobreaktrue
  \everypar{\if@nobreak
              \global\@nobreakfalse
              \clubpenalty \@M
              {\setbox\z@\lastbox}%
              \LastMac=\Nulle\relax%
            \else
              \clubpenalty \@clubpen
              \everypar{}%
            \fi}%
}

\newcount\gds@cbrk   \gds@cbrk=-300

\def\@nohdbrk{\interlinepenalty \@M\relax}

\let\@par=\par
\def\@restorepar{\def\par{\@par}}

\newif\if@endpe   \@endpefalse
 
\def\@doendpe{\@endpetrue \@nobreakfalse \LastMac=\Nulle\relax%
     \def\par{\@restorepar\everypar{}\par\@endpefalse}%
              \everypar{\setbox\z@\lastbox\everypar{}\@endpefalse}%
}

\def\section{\@ifstar{\@ssection}{\@section}}

\def\@section#1{
  \if@nobreak
    \everypar{}%
    \ifnum\LastMac=\Hae \addvspace{\half}\fi
  \else
    \addpen{\gds@cbrk}%
    \addvspace{\two}%
  \fi
  \bgroup
    \ninepoint\bf
    \Raggedright
    \global\advance\Sec \@ne
    \ifappendix
      \global\Eqnno=\z@ \global\SubEqnno=\z@\relax
      \def\ch@ck{#1}%
      \ifx\ch@ck\empty \def\c@lon{}\else\def\c@lon{:}\fi
      \noindent\@nohdbrk APPENDIX\ \thesection\c@lon\hskip 0.5em%
        \uppercase{#1}\par
    \else
      \noindent\@nohdbrk\thesection\hskip 1pc \uppercase{#1}\par
    \fi
    \global\SecSec=\z@
  \egroup
  \nobreak
  \vskip\half
  \nobreak
  \@noafterindent
  \LastMac=\Hae\relax
}

\def\@ssection#1{
  \if@nobreak
    \everypar{}%
    \ifnum\LastMac=\Hae \addvspace{\half}\fi
  \else
    \addpen{\gds@cbrk}%
    \addvspace{\two}%
  \fi
  \bgroup
    \ninepoint\bf
    \Raggedright
    \noindent\@nohdbrk\uppercase{#1}\par
  \egroup
  \nobreak
  \vskip\half
  \nobreak
  \@noafterindent
  \LastMac=\Hae\relax
}

\def\subsection{\@ifstar{\@ssubsection}{\@subsection}}

\def\@subsection#1{
  \if@nobreak
    \everypar{}%
    \ifnum\LastMac=\Hae \addvspace{1pt plus 1pt minus .5pt}\fi
  \else
    \addpen{\gds@cbrk}%
    \addvspace{\onehalf}%
  \fi
  \bgroup
    \ninepoint\bf
    \Raggedright
    \global\advance\SecSec \@ne
    \noindent\@nohdbrk\thesubsection \hskip 1pc\relax #1\par
    \global\SecSecSec=\z@
  \egroup
  \nobreak
  \vskip\half
  \nobreak
  \@noafterindent
  \LastMac=\Hbe\relax
}

\def\@ssubsection#1{
  \if@nobreak
    \everypar{}%
    \ifnum\LastMac=\Hae \addvspace{1pt plus 1pt minus .5pt}\fi
  \else
    \addpen{\gds@cbrk}%
    \addvspace{\onehalf}%
  \fi
  \bgroup
    \ninepoint\bf
    \Raggedright
    \noindent\@nohdbrk #1\par
  \egroup
  \nobreak
  \vskip\half
  \nobreak
  \@noafterindent
  \LastMac=\Hbe\relax
}

\def\subsubsection{\@ifstar{\@ssubsubsection}{\@subsubsection}}

\def\@subsubsection#1{
  \if@nobreak
    \everypar{}%
    \ifnum\LastMac=\Hbe \addvspace{1pt plus 1pt minus .5pt}\fi
  \else
    \addpen{\gds@cbrk}%
    \addvspace{\onehalf}%
  \fi
  \bgroup
    \ninepoint\it
    \Raggedright
    \global\advance\SecSecSec \@ne
    \noindent\@nohdbrk\thesubsubsection \hskip 1pc\relax #1\par
  \egroup
  \nobreak
  \vskip\half
  \nobreak
  \@noafterindent
  \LastMac=\Hce\relax
}

\def\@ssubsubsection#1{
  \if@nobreak
    \everypar{}%
    \ifnum\LastMac=\Hbe \addvspace{1pt plus 1pt minus .5pt}\fi
  \else
    \addpen{\gds@cbrk}%
    \addvspace{\onehalf}%
  \fi
  \bgroup
    \ninepoint\it
    \Raggedright
    \noindent\@nohdbrk #1\par
  \egroup
  \nobreak
  \vskip\half
  \nobreak
  \@noafterindent
  \LastMac=\Hce\relax
}

\def\paragraph#1{
  \if@nobreak
    \everypar{}%
  \else
    \addpen{\gds@cbrk}%
    \addvspace{\one}%
  \fi%
  \bgroup%
    \ninepoint\it
    \noindent #1\ \nobreak%
  \egroup
  \LastMac=\Hde\relax
  \ignorespaces
}


\newif\ifappendix

\def\appendix{%
  \global\appendixtrue
  \def\thesection{\Alph{Sec}}%
  \def\thesubsection{\thesection\arabic{SecSec}}%
  \def\theeq{\thesection\arabic{Eqnno}}%
  \Sec=\z@ \SecSec=\z@ \SecSecSec=\z@ \Eqnno=\z@ \SubEqnno=\z@\relax
}




\def\beginlist{%
  \par\if@nobreak \else\addvspace{\half}\fi%
  \bgroup%
    \ninepoint
    \let\item=\list@item%
}

\def\list@item{%
  \par\noindent\hskip 1em\relax%
  \ignorespaces%
}

\def\endlist{\par\egroup\addvspace{\half}\@doendpe}


\def\beginrefs{%
  \par
  \bgroup
    \eightpoint
    \Fullout
    \let\bibitem=\bib@item
}

\def\bib@item{%
  \par\parindent=1.5em\Hang{1.5em}{1}%
  \everypar={\Hang{1.5em}{1}\ignorespaces}%
  \noindent\ignorespaces
}

\def\endrefs{\par\egroup\@doendpe}


\newtoks\CatchLine

\def\@journal{Mon.\ Not.\ R.\ Astron.\ Soc.\ }  
\def\@pubyear{1994}        
\def\@pagerange{000--000}  
\def\@volume{000}          
\def\@microfiche{}         %

\def\pubyear#1{\gdef\@pubyear{#1}\@makecatchline}
\def\pagerange#1{\gdef\@pagerange{#1}\@makecatchline}
\def\volume#1{\gdef\@volume{#1}\@makecatchline}
\def\microfiche#1{\gdef\@microfiche{and Microfiche\ #1}\@makecatchline}

\def\@makecatchline{%
  \global\CatchLine{%
    {\rm \@journal {\bf \@volume},\ \@pagerange\ (\@pubyear)\ \@microfiche}}%
}

\@makecatchline 

\newtoks\LeftHeader
\def\shortauthor#1{
  \global\LeftHeader{#1}%
}

\newtoks\RightHeader
\def\shorttitle#1{
  \global\RightHeader{#1}%
}

\def\PageHead{
  \begingroup
    \ifsp@page
      \csname ps@\sp@type\endcsname
    \fi
    \ifodd\pageno
      \let\the@head=\@oddhead
    \else
      \let\the@head=\@evenhead
    \fi
    \vbox to \z@{\vskip-22.5\p@%
      \hbox to \PageWidth{\vbox to8.5\p@{}%
        \the@head
      }%
    \vss}%
  \endgroup
  \nointerlineskip
}

\gdef\PageFoot{%
  \nointerlineskip%
  \begingroup
  \ifsp@page
    \csname ps@\sp@type\endcsname
    \global\sp@pagefalse
  \fi
  \vbox to 22pt{\vfil%
    \hbox to \PageWidth{%
      \eightpoint\strut\noindent
      \ifodd\pageno
        \@oddfoot
      \else
        \@evenfoot
      \fi
    }%
  }%
  \endgroup
}

\def\today{%
  \number\day\space
  \ifcase\month\or January\or February\or March\or April\or May\or June\or
    July\or August\or September\or October\or November\or December\fi
  \space\number\year%
}

\def\authorcomment#1{%
  \gdef\PageFoot{%
    \nointerlineskip%
    \vbox to 20pt{\vfil%
      \hbox to \PageWidth{\elevenpoint\noindent \hfil #1 \hfil}}%
  }%
}


\newif\ifplate@page
\newbox\plt@box

\def\beginplatepage{%
  \let\plate=\plate@head
  \let\caption=\fig@caption
  \global\setbox\plt@box=\vbox\bgroup
  \TEMPDIMEN=\PageWidth 
  \hsize=\PageWidth\relax
}

\def\endplatepage{\par\egroup\global\plate@pagetrue}
\def\plate@head#1{\gdef\plt@cap{#1}}


\def\letters{%
  \gdef\folio{\ifnum\pageno<\z@ L\romannumeral-\pageno
    \else L\number\pageno \fi}%
}


\newdimen\mathindent

\global\mathindent=\z@
\global\everydisplay{\global\@dspwd=\displaywidth\displaysetup}


\def\@displaylines#1{
  {}$\displ@y\hbox{\vbox{\halign{$\@lign\hfil\displaystyle##\hfil$\crcr
  #1\crcr}}}${}%
}

\def\@eqalign#1{\null\vcenter{\openup\jot\m@th
  \ialign{\strut\hfil$\displaystyle{##}$&$\displaystyle{{}##}$\hfil
      \crcr#1\crcr}}%
}

\def\@eqalignno#1{
  \global\advance\@dspwd by -\mathindent%
  {}$\displ@y\hbox{\vbox{\halign to\@dspwd%
  {\hfil$\@lign\displaystyle{##}$\tabskip\z@skip
  &$\@lign\displaystyle{{}##}$\hfil\tabskip\centering
  &\llap{$\@lign##$}\tabskip\z@skip\crcr
  #1\crcr}}}${}%
}


\global\let\displaylines=\@displaylines
\global\let\eqalign=\@eqalign
\global\let\eqalignno=\@eqalignno
\global\let\leqalignno=\@eqalignno

\newdimen\@dspwd   \@dspwd=\z@
\newif\if@eqno
\newif\if@leqno
\newtoks\@eqn
\newtoks\@eq

\def\displaysetup#1$${\displaytest#1\eqno\eqno\displaytest}

\def\displaytest#1\eqno#2\eqno#3\displaytest{%
 \if!#3!\ldisplaytest#1\leqno\leqno\ldisplaytest
 \else\@eqnotrue\@leqnofalse\@eqn={#2}\@eq={#1}\fi
 \generaldisplay$$}

\def\ldisplaytest#1\leqno#2\leqno#3\ldisplaytest{%
\@eq={#1}%
 \if!#3!\@eqnofalse\else\@eqnotrue\@leqnotrue
  \@eqn={#2}\fi}

\def\generaldisplay{%
  \if@eqno
    \if@leqno
      \hbox to \displaywidth{\noindent
        \rlap{$\displaystyle\the\@eqn$}%
        \hskip\mathindent$\displaystyle\the\@eq$\hfil}%
    \else
      \hbox to \displaywidth{\noindent
        \hskip\mathindent
        $\displaystyle\the\@eq$\hfil$\displaystyle\the\@eqn$}%
    \fi
  \else
    \hbox to \displaywidth{\noindent
      \hskip\mathindent$\displaystyle\the\@eq$\hfil}%
  \fi
}


\def\@notice{%
  \par\Two%
  \noindent{\b@ls{11pt}\ninerm This paper has been produced using the
    Royal Astronomical Society/Blackwell Science \TeX\ macros.\par}%
}

\outer\def\bye{\@notice\par\vfill\supereject\end}


\def\start@mess{%
  Monthly notices of the RAS journal style (\@typeface)\space
    v\@version,\space \@verdate.%
}

\everyjob{\Warn{\start@mess}}



\newif\if@debug \@debugfalse  

\def\Print#1{\if@debug\immediate\write16{#1}\else \fi}
\def\Warn#1{\immediate\write16{#1}}
\def\wlog#1{}

\newcount\Iteration 

\def\Single{0} \def\Double{1}                 
\def\Figure{0} \def\Table{1}                  

\def\InStack{0}  
\def\InZoneA{1}
\def\InZoneB{2}
\def\InZoneC{3}

\newcount\TEMPCOUNT 
\newdimen\TEMPDIMEN 
\newbox\TEMPBOX     
\newbox\VOIDBOX     

\newcount\LengthOfStack 
\newcount\MaxItems      
\newcount\StackPointer
\newcount\Point         
\newcount\NextFigure    
\newcount\NextTable     
\newcount\NextItem      

\newcount\StatusStack   
\newcount\NumStack      
\newcount\TypeStack     
\newcount\SpanStack     
\newcount\BoxStack      

\newcount\ItemSTATUS    
\newcount\ItemNUMBER    
\newcount\ItemTYPE      
\newcount\ItemSPAN      
\newbox\ItemBOX         
\newdimen\ItemSIZE      

\newdimen\PageHeight    
\newdimen\TextLeading   
\newdimen\Feathering    
\newcount\LinesPerPage  
\newdimen\ColumnWidth   
\newdimen\ColumnGap     
\newdimen\PageWidth     
\newdimen\BodgeHeight   
\newcount\Leading       

\newdimen\ZoneBSize  
\newdimen\TextSize   
\newbox\ZoneABOX     
\newbox\ZoneBBOX     
\newbox\ZoneCBOX     

\newif\ifFirstSingleItem
\newif\ifFirstZoneA
\newif\ifMakePageInComplete
\newif\ifMoreFigures \MoreFiguresfalse 
\newif\ifMoreTables  \MoreTablesfalse  

\newif\ifFigInZoneB 
\newif\ifFigInZoneC 
\newif\ifTabInZoneB 
\newif\ifTabInZoneC

\newif\ifZoneAFullPage

\newbox\MidBOX    
\newbox\LeftBOX
\newbox\RightBOX
\newbox\PageBOX   

\newif\ifLeftCOL  
\LeftCOLtrue

\newdimen\ZoneBAdjust

\newcount\ItemFits
\def\Yes{1}
\def\No{2}


\MaxItems=15
\NextFigure=\z@        
\NextTable=\@ne

\BodgeHeight=6pt
\TextLeading=11pt    
\Leading=11
\Feathering=\z@      
\LinesPerPage=61     
\topskip=\TextLeading
\ColumnWidth=20pc    
\ColumnGap=2pc       

\newskip\ItemSepamount  
\ItemSepamount=\TextLeading plus \TextLeading minus 4pt

\parskip=\z@ plus .1pt
\parindent=18pt
\widowpenalty=\z@
\clubpenalty=10000
\tolerance=1500
\hbadness=1500
\abovedisplayskip=6pt plus 2pt minus 1pt
\belowdisplayskip=6pt plus 2pt minus 1pt
\abovedisplayshortskip=6pt plus 2pt minus 1pt
\belowdisplayshortskip=6pt plus 2pt minus 1pt

\frenchspacing

\ninepoint 

\PageHeight=682pt
\PageWidth=2\ColumnWidth
\advance\PageWidth by \ColumnGap

\pagestyle{headings}




\newcount\DUMMY \StatusStack=\allocationnumber
\newcount\DUMMY \newcount\DUMMY \newcount\DUMMY 
\newcount\DUMMY \newcount\DUMMY \newcount\DUMMY 
\newcount\DUMMY \newcount\DUMMY \newcount\DUMMY
\newcount\DUMMY \newcount\DUMMY \newcount\DUMMY 
\newcount\DUMMY \newcount\DUMMY \newcount\DUMMY

\newcount\DUMMY \NumStack=\allocationnumber
\newcount\DUMMY \newcount\DUMMY \newcount\DUMMY 
\newcount\DUMMY \newcount\DUMMY \newcount\DUMMY 
\newcount\DUMMY \newcount\DUMMY \newcount\DUMMY 
\newcount\DUMMY \newcount\DUMMY \newcount\DUMMY 
\newcount\DUMMY \newcount\DUMMY \newcount\DUMMY

\newcount\DUMMY \TypeStack=\allocationnumber
\newcount\DUMMY \newcount\DUMMY \newcount\DUMMY 
\newcount\DUMMY \newcount\DUMMY \newcount\DUMMY 
\newcount\DUMMY \newcount\DUMMY \newcount\DUMMY 
\newcount\DUMMY \newcount\DUMMY \newcount\DUMMY 
\newcount\DUMMY \newcount\DUMMY \newcount\DUMMY

\newcount\DUMMY \SpanStack=\allocationnumber
\newcount\DUMMY \newcount\DUMMY \newcount\DUMMY 
\newcount\DUMMY \newcount\DUMMY \newcount\DUMMY 
\newcount\DUMMY \newcount\DUMMY \newcount\DUMMY 
\newcount\DUMMY \newcount\DUMMY \newcount\DUMMY 
\newcount\DUMMY \newcount\DUMMY \newcount\DUMMY

\newbox\DUMMY   \BoxStack=\allocationnumber
\newbox\DUMMY   \newbox\DUMMY \newbox\DUMMY 
\newbox\DUMMY   \newbox\DUMMY \newbox\DUMMY 
\newbox\DUMMY   \newbox\DUMMY \newbox\DUMMY 
\newbox\DUMMY   \newbox\DUMMY \newbox\DUMMY 
\newbox\DUMMY   \newbox\DUMMY \newbox\DUMMY

\def\wlog{\immediate\write\m@ne}


\def\GetItemAll#1{%
 \GetItemSTATUS{#1}
 \GetItemNUMBER{#1}
 \GetItemTYPE{#1}
 \GetItemSPAN{#1}
 \GetItemBOX{#1}
}

\def\GetItemSTATUS#1{%
 \Point=\StatusStack
 \advance\Point by #1
 \global\ItemSTATUS=\count\Point
}

\def\GetItemNUMBER#1{%
 \Point=\NumStack
 \advance\Point by #1
 \global\ItemNUMBER=\count\Point
}

\def\GetItemTYPE#1{%
 \Point=\TypeStack
 \advance\Point by #1
 \global\ItemTYPE=\count\Point
}

\def\GetItemSPAN#1{%
 \Point\SpanStack
 \advance\Point by #1
 \global\ItemSPAN=\count\Point
}

\def\GetItemBOX#1{%
 \Point=\BoxStack
 \advance\Point by #1
 \global\setbox\ItemBOX=\vbox{\copy\Point}
 \global\ItemSIZE=\ht\ItemBOX
 \global\advance\ItemSIZE by \dp\ItemBOX
 \TEMPCOUNT=\ItemSIZE
 \divide\TEMPCOUNT by \Leading
 \divide\TEMPCOUNT by 65536
 \advance\TEMPCOUNT \@ne
 \ItemSIZE=\TEMPCOUNT pt
 \global\multiply\ItemSIZE by \Leading
}


\def\JoinStack{%
 \ifnum\LengthOfStack=\MaxItems 
  \Warn{WARNING: Stack is full...some items will be lost!}
 \else
  \Point=\StatusStack
  \advance\Point by \LengthOfStack
  \global\count\Point=\ItemSTATUS
  \Point=\NumStack
  \advance\Point by \LengthOfStack
  \global\count\Point=\ItemNUMBER
  \Point=\TypeStack
  \advance\Point by \LengthOfStack
  \global\count\Point=\ItemTYPE
  \Point\SpanStack
  \advance\Point by \LengthOfStack
  \global\count\Point=\ItemSPAN
  \Point=\BoxStack
  \advance\Point by \LengthOfStack
  \global\setbox\Point=\vbox{\copy\ItemBOX}
  \global\advance\LengthOfStack \@ne
  \ifnum\ItemTYPE=\Figure 
   \global\MoreFigurestrue
  \else
   \global\MoreTablestrue
  \fi
 \fi
}


\def\LeaveStack#1{%
 {\Iteration=#1
 \loop
 \ifnum\Iteration<\LengthOfStack
  \advance\Iteration \@ne
  \GetItemSTATUS{\Iteration}
   \advance\Point by \m@ne
   \global\count\Point=\ItemSTATUS
  \GetItemNUMBER{\Iteration}
   \advance\Point by \m@ne
   \global\count\Point=\ItemNUMBER
  \GetItemTYPE{\Iteration}
   \advance\Point by \m@ne
   \global\count\Point=\ItemTYPE
  \GetItemSPAN{\Iteration}
   \advance\Point by \m@ne
   \global\count\Point=\ItemSPAN
  \GetItemBOX{\Iteration}
   \advance\Point by \m@ne
   \global\setbox\Point=\vbox{\copy\ItemBOX}
 \repeat}
 \global\advance\LengthOfStack by \m@ne
}


\newif\ifStackNotClean

\def\CleanStack{%
 \StackNotCleantrue
 {\Iteration=\z@
  \loop
   \ifStackNotClean
    \GetItemSTATUS{\Iteration}
    \ifnum\ItemSTATUS=\InStack
     \advance\Iteration \@ne
     \else
      \LeaveStack{\Iteration}
    \fi
   \ifnum\LengthOfStack<\Iteration
    \StackNotCleanfalse
   \fi
 \repeat}
}


\def\FindItem#1#2{%
 \global\StackPointer=\m@ne 
 {\Iteration=\z@
  \loop
  \ifnum\Iteration<\LengthOfStack
   \GetItemSTATUS{\Iteration}
   \ifnum\ItemSTATUS=\InStack
    \GetItemTYPE{\Iteration}
    \ifnum\ItemTYPE=#1
     \GetItemNUMBER{\Iteration}
     \ifnum\ItemNUMBER=#2
      \global\StackPointer=\Iteration
      \Iteration=\LengthOfStack 
     \fi
    \fi
   \fi
  \advance\Iteration \@ne
 \repeat}
}


\def\FindNext{%
 \global\StackPointer=\m@ne 
 {\Iteration=\z@
  \loop
  \ifnum\Iteration<\LengthOfStack
   \GetItemSTATUS{\Iteration}
   \ifnum\ItemSTATUS=\InStack
    \GetItemTYPE{\Iteration}
   \ifnum\ItemTYPE=\Figure
    \ifMoreFigures
      \global\NextItem=\Figure
      \global\StackPointer=\Iteration
      \Iteration=\LengthOfStack 
    \fi
   \fi
   \ifnum\ItemTYPE=\Table
    \ifMoreTables
      \global\NextItem=\Table
      \global\StackPointer=\Iteration
      \Iteration=\LengthOfStack 
    \fi
   \fi
  \fi
  \advance\Iteration \@ne
 \repeat}
}


\def\ChangeStatus#1#2{%
 \Point=\StatusStack
 \advance\Point by #1
 \global\count\Point=#2
}



\def\Zone{\InZoneA}

\ZoneBAdjust=\z@

\def\MakePage{
 \global\ZoneBSize=\PageHeight
 \global\TextSize=\ZoneBSize
 \global\ZoneAFullPagefalse
 \global\topskip=\TextLeading
 \MakePageInCompletetrue
 \MoreFigurestrue
 \MoreTablestrue
 \FigInZoneBfalse
 \FigInZoneCfalse
 \TabInZoneBfalse
 \TabInZoneCfalse
 \global\FirstSingleItemtrue
 \global\FirstZoneAtrue
 \global\setbox\ZoneABOX=\box\VOIDBOX
 \global\setbox\ZoneBBOX=\box\VOIDBOX
 \global\setbox\ZoneCBOX=\box\VOIDBOX
 \loop
  \ifMakePageInComplete
 \FindNext
 \ifnum\StackPointer=\m@ne
  \NextItem=\m@ne
  \MoreFiguresfalse
  \MoreTablesfalse
 \fi
 \ifnum\NextItem=\Figure
   \FindItem{\Figure}{\NextFigure}
   \ifnum\StackPointer=\m@ne \global\MoreFiguresfalse
   \else
    \GetItemSPAN{\StackPointer}
    \ifnum\ItemSPAN=\Single \def\Zone{\InZoneB}\relax
     \ifFigInZoneC \global\MoreFiguresfalse\fi
    \else
     \def\Zone{\InZoneA}
     \ifFigInZoneB \def\Zone{\InZoneC}\fi
    \fi
   \fi
   \ifMoreFigures\Print{}\FigureItems\fi
 \fi
\ifnum\NextItem=\Table
   \FindItem{\Table}{\NextTable}
   \ifnum\StackPointer=\m@ne \global\MoreTablesfalse
   \else
    \GetItemSPAN{\StackPointer}
    \ifnum\ItemSPAN=\Single\relax
     \ifTabInZoneC \global\MoreTablesfalse\fi
    \else
     \def\Zone{\InZoneA}
     \ifTabInZoneB \def\Zone{\InZoneC}\fi
    \fi
   \fi
   \ifMoreTables\Print{}\TableItems\fi
 \fi
   \MakePageInCompletefalse 
   \ifMoreFigures\MakePageInCompletetrue\fi
   \ifMoreTables\MakePageInCompletetrue\fi
 \repeat
 \ifZoneAFullPage
  \global\TextSize=\z@
  \global\ZoneBSize=\z@
  \global\vsize=\z@\relax
  \global\topskip=\z@\relax
  \vbox to \z@{\vss}
  \eject
 \else
 \global\advance\ZoneBSize by -\ZoneBAdjust
 \global\vsize=\ZoneBSize
 \global\hsize=\ColumnWidth
 \global\ZoneBAdjust=\z@
 \ifdim\TextSize<23pt
 \Warn{}
 \Warn{* Making column fall short: TextSize=\the\TextSize *}
 \vskip-\lastskip\eject\fi
 \fi
}

\def\MakeRightCol{
 \global\TextSize=\ZoneBSize
 \MakePageInCompletetrue
 \MoreFigurestrue
 \MoreTablestrue
 \global\FirstSingleItemtrue
 \global\setbox\ZoneBBOX=\box\VOIDBOX
 \def\Zone{\InZoneB}
 \loop
  \ifMakePageInComplete
 \FindNext
 \ifnum\StackPointer=\m@ne
  \NextItem=\m@ne
  \MoreFiguresfalse
  \MoreTablesfalse
 \fi
 \ifnum\NextItem=\Figure
   \FindItem{\Figure}{\NextFigure}
   \ifnum\StackPointer=\m@ne \MoreFiguresfalse
   \else
    \GetItemSPAN{\StackPointer}
    \ifnum\ItemSPAN=\Double\relax
     \MoreFiguresfalse\fi
   \fi
   \ifMoreFigures\Print{}\FigureItems\fi
 \fi
 \ifnum\NextItem=\Table
   \FindItem{\Table}{\NextTable}
   \ifnum\StackPointer=\m@ne \MoreTablesfalse
   \else
    \GetItemSPAN{\StackPointer}
    \ifnum\ItemSPAN=\Double\relax
     \MoreTablesfalse\fi
   \fi
   \ifMoreTables\Print{}\TableItems\fi
 \fi
   \MakePageInCompletefalse 
   \ifMoreFigures\MakePageInCompletetrue\fi
   \ifMoreTables\MakePageInCompletetrue\fi
 \repeat
 \ifZoneAFullPage
  \global\TextSize=\z@
  \global\ZoneBSize=\z@
  \global\vsize=\z@\relax
  \global\topskip=\z@\relax
  \vbox to \z@{\vss}
  \eject
 \else
 \global\vsize=\ZoneBSize
 \global\hsize=\ColumnWidth
 \ifdim\TextSize<23pt
 \Warn{}
 \Warn{* Making column fall short: TextSize=\the\TextSize *}
 \vskip-\lastskip\eject\fi
\fi
}

\def\FigureItems{
 \Print{Considering...}
 \ShowItem{\StackPointer}
 \GetItemBOX{\StackPointer} 
 \GetItemSPAN{\StackPointer}
  \CheckFitInZone 
  \ifnum\ItemFits=\Yes
   \ifnum\ItemSPAN=\Single
     \ChangeStatus{\StackPointer}{\InZoneB} 
     \global\FigInZoneBtrue
     \ifFirstSingleItem
      \hbox{}\vskip-\BodgeHeight
     \global\advance\ItemSIZE by \TextLeading
     \fi
     \unvbox\ItemBOX\ItemSep
     \global\FirstSingleItemfalse
     \global\advance\TextSize by -\ItemSIZE
     \global\advance\TextSize by -\TextLeading
   \else
    \ifFirstZoneA
     \global\advance\ItemSIZE by \TextLeading
     \global\FirstZoneAfalse\fi
    \global\advance\TextSize by -\ItemSIZE
    \global\advance\TextSize by -\TextLeading
    \global\advance\ZoneBSize by -\ItemSIZE
    \global\advance\ZoneBSize by -\TextLeading
    \ifFigInZoneB\relax
     \else
     \ifdim\TextSize<3\TextLeading
     \global\ZoneAFullPagetrue
     \fi
    \fi
    \ChangeStatus{\StackPointer}{\Zone}
    \ifnum\Zone=\InZoneC \global\FigInZoneCtrue\fi
  \fi
   \Print{TextSize=\the\TextSize}
   \Print{ZoneBSize=\the\ZoneBSize}
  \global\advance\NextFigure \@ne
   \Print{This figure has been placed.}
  \else
   \Print{No space available for this figure...holding over.}
   \Print{}
   \global\MoreFiguresfalse
  \fi
}

\def\TableItems{
 \Print{Considering...}
 \ShowItem{\StackPointer}
 \GetItemBOX{\StackPointer} 
 \GetItemSPAN{\StackPointer}
  \CheckFitInZone 
  \ifnum\ItemFits=\Yes
   \ifnum\ItemSPAN=\Single
    \ChangeStatus{\StackPointer}{\InZoneB}
     \global\TabInZoneBtrue
     \ifFirstSingleItem
      \hbox{}\vskip-\BodgeHeight
     \global\advance\ItemSIZE by \TextLeading
     \fi
     \unvbox\ItemBOX\ItemSep
     \global\FirstSingleItemfalse
     \global\advance\TextSize by -\ItemSIZE
     \global\advance\TextSize by -\TextLeading
   \else
    \ifFirstZoneA
    \global\advance\ItemSIZE by \TextLeading
    \global\FirstZoneAfalse\fi
    \global\advance\TextSize by -\ItemSIZE
    \global\advance\TextSize by -\TextLeading
    \global\advance\ZoneBSize by -\ItemSIZE
    \global\advance\ZoneBSize by -\TextLeading
    \ifFigInZoneB\relax
     \else
     \ifdim\TextSize<3\TextLeading
     \global\ZoneAFullPagetrue
     \fi
    \fi
    \ChangeStatus{\StackPointer}{\Zone}
    \ifnum\Zone=\InZoneC \global\TabInZoneCtrue\fi
   \fi
  \global\advance\NextTable \@ne
   \Print{This table has been placed.}
  \else
  \Print{No space available for this table...holding over.}
   \Print{}
   \global\MoreTablesfalse
  \fi
}


\def\CheckFitInZone{%
{\advance\TextSize by -\ItemSIZE
 \advance\TextSize by -\TextLeading
 \ifFirstSingleItem
  \advance\TextSize by \TextLeading
 \fi
 \ifnum\Zone=\InZoneA\relax
  \else \advance\TextSize by -\ZoneBAdjust
 \fi
 \ifdim\TextSize<3\TextLeading \global\ItemFits=\No
 \else \global\ItemFits=\Yes\fi}
}

\def\BeginOpening{%
  \ninepoint
  \thispagestyle{titlepage}%
  \global\setbox\ItemBOX=\vbox\bgroup%
    \hsize=\PageWidth%
    \hrule height \z@
    \ifsinglecol\vskip 6pt\fi 
}

\let\begintopmatter=\BeginOpening  

\def\EndOpening{%
  \One
  \egroup
  \ifsinglecol
    \box\ItemBOX%
    \vskip\TextLeading plus 2\TextLeading
    \@noafterindent
  \else
    \ItemNUMBER=\z@%
    \ItemTYPE=\Figure
    \ItemSPAN=\Double
    \ItemSTATUS=\InStack
    \JoinStack
  \fi
}


\newif\if@here  \@herefalse

\def\no@float{\global\@heretrue}
\let\nofloat=\relax 

\def\beginfigure{%
  \@ifstar{\global\@dfloattrue \@bfigure}{\global\@dfloatfalse \@bfigure}%
}

\def\@bfigure#1{%
  \par
  \if@dfloat
    \ItemSPAN=\Double
    \TEMPDIMEN=\PageWidth
  \else
    \ItemSPAN=\Single
    \TEMPDIMEN=\ColumnWidth
  \fi
  \ifsinglecol
    \TEMPDIMEN=\PageWidth
  \else
    \ItemSTATUS=\InStack
    \ItemNUMBER=#1%
    \ItemTYPE=\Figure
  \fi
  \bgroup
    \hsize=\TEMPDIMEN
    \global\setbox\ItemBOX=\vbox\bgroup
      \eightpoint\nostb@ls{10pt}%
      \let\caption=\fig@caption
      \ifsinglecol \let\nofloat=\no@float\fi
}

\def\fig@caption#1{%
  \vskip 5.5pt plus 6pt%
  \bgroup 
    \eightpoint\nostb@ls{10pt}%
    \setbox\TEMPBOX=\hbox{#1}%
    \ifdim\wd\TEMPBOX>\TEMPDIMEN
      \noindent \unhbox\TEMPBOX\par
    \else
      \hbox to \hsize{\hfil\unhbox\TEMPBOX\hfil}%
    \fi
  \egroup
}

\def\endfigure{%
  \par\egroup 
  \egroup
  \ifsinglecol
    \if@here \midinsert\global\@herefalse\else \topinsert\fi
      \unvbox\ItemBOX
    \endinsert
  \else
    \JoinStack
    \Print{Processing source for figure \the\ItemNUMBER}%
  \fi
}


\newbox\tab@cap@box
\def\tab@caption#1{\global\setbox\tab@cap@box=\hbox{#1\par}}

\newtoks\tab@txt@toks
\long\def\tab@txt#1{\global\tab@txt@toks={#1}\global\table@txttrue}

\newif\iftable@txt  \table@txtfalse
\newif\if@dfloat    \@dfloatfalse

\def\begintable{%
  \@ifstar{\global\@dfloattrue \@btable}{\global\@dfloatfalse \@btable}%
}

\def\@btable#1{%
  \par
  \if@dfloat
    \ItemSPAN=\Double
    \TEMPDIMEN=\PageWidth
  \else
    \ItemSPAN=\Single
    \TEMPDIMEN=\ColumnWidth
  \fi
  \ifsinglecol
    \TEMPDIMEN=\PageWidth
  \else
    \ItemSTATUS=\InStack
    \ItemNUMBER=#1%
    \ItemTYPE=\Table
  \fi
  \bgroup
    \eightpoint\nostb@ls{10pt}%
    \global\setbox\ItemBOX=\vbox\bgroup
      \let\caption=\tab@caption
      \let\tabletext=\tab@txt
      \ifsinglecol \let\nofloat=\no@float\fi
}

\def\endtable{%
  \par\egroup 
  \egroup
  \setbox\TEMPBOX=\hbox to \TEMPDIMEN{%
    \eightpoint\nostb@ls{10pt}%
    \hss
    \vbox{%
      \hsize=\wd\ItemBOX
      \ifvoid\tab@cap@box
      \else
        \noindent\unhbox\tab@cap@box
        \vskip 5.5pt plus 6pt%
      \fi
      \box\ItemBOX
      \iftable@txt
        \vskip 10pt%
        \noindent\the\tab@txt@toks
        \global\table@txtfalse
      \fi
    }%
    \hss
  }%
  \ifsinglecol
    \if@here \midinsert\global\@herefalse\else \topinsert\fi
      \box\TEMPBOX
    \endinsert
  \else
    \global\setbox\ItemBOX=\box\TEMPBOX
    \JoinStack
    \Print{Processing source for table \the\ItemNUMBER}%
  \fi
}

\def\UnloadZoneA{%
\FirstZoneAtrue
 \Iteration=\z@
  \loop
   \ifnum\Iteration<\LengthOfStack
    \GetItemSTATUS{\Iteration}
    \ifnum\ItemSTATUS=\InZoneA
     \GetItemBOX{\Iteration}
     \ifFirstZoneA \vbox to \BodgeHeight{\vfil}%
     \FirstZoneAfalse\fi
     \unvbox\ItemBOX\ItemSep
     \LeaveStack{\Iteration}
     \else
     \advance\Iteration \@ne
   \fi
 \repeat
}

\def\UnloadZoneC{%
\Iteration=\z@
  \loop
   \ifnum\Iteration<\LengthOfStack
    \GetItemSTATUS{\Iteration}
    \ifnum\ItemSTATUS=\InZoneC
     \GetItemBOX{\Iteration}
     \ItemSep\unvbox\ItemBOX
     \LeaveStack{\Iteration}
     \else
     \advance\Iteration \@ne
   \fi
 \repeat
}


\def\ShowItem#1{
  {\GetItemAll{#1}
  \Print{\the#1:
  {TYPE=\ifnum\ItemTYPE=\Figure Figure\else Table\fi}
  {NUMBER=\the\ItemNUMBER}
  {SPAN=\ifnum\ItemSPAN=\Single Single\else Double\fi}
  {SIZE=\the\ItemSIZE}}}
}

\def\ShowStack{%
 \Print{}
 \Print{LengthOfStack = \the\LengthOfStack}
 \ifnum\LengthOfStack=\z@ \Print{Stack is empty}\fi
 \Iteration=\z@
 \loop
 \ifnum\Iteration<\LengthOfStack
  \ShowItem{\Iteration}
  \advance\Iteration \@ne
 \repeat
}

\def\B#1#2{%
\hbox{\vrule\kern-0.4pt\vbox to #2{%
\hrule width #1\vfill\hrule}\kern-0.4pt\vrule}
}


\newif\ifsinglecol   \singlecolfalse

\def\onecolumn{%
  \global\output={\singlecoloutput}%
  \global\hsize=\PageWidth
  \global\vsize=\PageHeight
  \global\ColumnWidth=\hsize
  \global\TextLeading=12pt
  \global\Leading=12
  \global\singlecoltrue
  \global\let\onecolumn=\relax
  \global\let\footnote=\sing@footnote
  \global\let\vfootnote=\sing@vfootnote
  \ninepoint 
  \message{(Single column)}%
}

\def\singlecoloutput{%
  \shipout\vbox{\PageHead\vbox to \PageHeight{\pagebody\vss}\PageFoot}%
  \advancepageno
  \ifplate@page
    \shipout\vbox{%
      \sp@pagetrue
      \def\sp@type{plate}%
      \global\plate@pagefalse
      \PageHead\vbox to \PageHeight{\unvbox\plt@box\vfil}\PageFoot%
    }%
    \message{[plate]}%
    \advancepageno
  \fi
  \ifnum\outputpenalty>-\@MM \else\dosupereject\fi%
}

\def\ItemSep{\vskip\ItemSepamount\relax}

\def\ItemSepbreak{\par\ifdim\lastskip<\ItemSepamount
  \removelastskip\penalty-200\ItemSep\fi%
}


\let\@@endinsert=\endinsert 

\def\endinsert{\egroup 
  \if@mid \dimen@\ht\z@ \advance\dimen@\dp\z@ \advance\dimen@12\p@
    \advance\dimen@\pagetotal \advance\dimen@-\pageshrink
    \ifdim\dimen@>\pagegoal\@midfalse\p@gefalse\fi\fi
  \if@mid \ItemSep\box\z@\ItemSepbreak
  \else\insert\topins{\penalty100 
    \splittopskip\z@skip
    \splitmaxdepth\maxdimen \floatingpenalty\z@
    \ifp@ge \dimen@\dp\z@
    \vbox to\vsize{\unvbox\z@\kern-\dimen@}
    \else \box\z@\nobreak\ItemSep\fi}\fi\endgroup%
}


\def\gobbleone#1{}
\def\gobbletwo#1#2{}
\let\footnote=\gobbletwo 
\let\vfootnote=\gobbleone

\def\sing@footnote#1{\let\@sf\empty 
  \ifhmode\edef\@sf{\spacefactor\the\spacefactor}\/\fi
  \hbox{$^{\hbox{\eightpoint #1}}$}\@sf\sing@vfootnote{#1}%
}

\def\sing@vfootnote#1{\insert\footins\bgroup\eightpoint\b@ls{9pt}%
  \interlinepenalty\interfootnotelinepenalty
  \splittopskip\ht\strutbox 
  \splitmaxdepth\dp\strutbox \floatingpenalty\@MM
  \leftskip\z@skip \rightskip\z@skip \spaceskip\z@skip \xspaceskip\z@skip
  \noindent $^{\scriptstyle\hbox{#1}}$\hskip 4pt%
    \footstrut\futurelet\next\fo@t%
}

\def\footnoterule{\kern-3\p@ \hrule height \z@ \kern 3\p@}

\skip\footins=19.5pt plus 12pt minus 1pt
\count\footins=1000
\dimen\footins=\maxdimen

\def\note#1#2{%
  \let\@sf=\empty \ifhmode\edef\@sf{\spacefactor\the\spacefactor}\/\fi
  #1\insert\footins\bgroup
    \eightpoint\b@ls{10pt}\rm
    \interlinepenalty\interfootnotelinepenalty
    \splitmaxdepth\dp\strutbox \floatingpenalty\@MM
    \leftskip\z@skip \rightskip\z@skip \spaceskip\z@skip \xspaceskip\z@skip
    \noindent\footstrut #1$\,$#2\strut\par
  \egroup
  \@sf\relax}


\def\landscape{%
  \global\TEMPDIMEN=\PageWidth
  \global\PageWidth=\PageHeight
  \global\PageHeight=\TEMPDIMEN
  \global\let\landscape=\relax
  \onecolumn
  \message{(landscape)}%
  \raggedbottom
}


\output{%
  \ifLeftCOL
    \global\setbox\LeftBOX=\vbox to \ZoneBSize{\box255\unvbox\ZoneBBOX
      \ifvoid\footins\else
        \vskip\skip\footins\unvbox\footins\fi
    }%
    \global\LeftCOLfalse
    \MakeRightCol
  \else
    \setbox\RightBOX=\vbox to \ZoneBSize{\box255\unvbox\ZoneBBOX
      \ifvoid\footins\else
        \vskip\skip\footins\unvbox\footins\fi
    }%
    \setbox\MidBOX=\hbox{\box\LeftBOX\hskip\ColumnGap\box\RightBOX}%
    \setbox\PageBOX=\vbox to \PageHeight{%
      \UnloadZoneA\box\MidBOX\UnloadZoneC}%
    \shipout\vbox{\PageHead\vbox to \PageHeight{\box\PageBOX\vss}\PageFoot}%
    \advancepageno
    \ifplate@page
      \shipout\vbox{%
        \sp@pagetrue
        \def\sp@type{plate}%
        \global\plate@pagefalse
        \PageHead\vbox to \PageHeight{\unvbox\plt@box\vfil}\PageFoot%
      }%
      \message{[plate]}%
      \advancepageno
    \fi
    \global\LeftCOLtrue
    \CleanStack
    \MakePage
  \fi
}


\Warn{\start@mess}

\newif\ifCUPmtplainloaded 
\ifprod@font
  \global\CUPmtplainloadedtrue
\fi

\def\mnmacrosloaded{} 

\catcode `\@=12 


\fi

%

\newif\ifAMStwofonts

\ifCUPmtplainloaded \else
  \NewTextAlphabet{textbfit} {cmbxti10} {}
  \NewTextAlphabet{textbfss} {cmssbx10} {}
  \NewMathAlphabet{mathbfit} {cmbxti10} {} 
  \NewMathAlphabet{mathbfss} {cmssbx10} {} 
  \ifAMStwofonts
    \NewSymbolFont{upmath} {eurm10}
    \NewSymbolFont{AMSa} {msam10}
    \NewMathSymbol{\upi}     {0}{upmath}{19}
    \NewMathSymbol{\umu}     {0}{upmath}{16}
    \NewMathSymbol{\upartial}{0}{upmath}{40}
    \NewMathSymbol{\leqslant}{3}{AMSa}{36}
    \NewMathSymbol{\geqslant}{3}{AMSa}{3E}

    \let\leq=\leqslant \let\le=\leqslant
    \let\geq=\geqslant 
  \else
    \def\umu{\mu}
    \def\upi{\pi}
    \def\upartial{\partial}
  \fi
\fi


\pageoffset{-2.5pc}{0pc}

\loadboldmathnames



\pagerange{}    
\pubyear{}
\volume{}

\begintopmatter  

\title {Non-linear evolution of thermally unstable slim  
 accretion discs with a diffusive form of viscosity}

\author{Ewa Szuszkiewicz$^{1,2,3,4}$ and John C. Miller$^{4,5}$}

\affiliation{$^1$Institute of Physics, University of Szczecin, 
ul. Wielkopolska 15, 70-451 Szczecin, Poland \break
$^2$Torun Centre for Astronomy, Nicolaus Copernicus 
University, ul. Gagarina 11, 87-100 Toru\'n, Poland \break 
$^3$Dipartimento di Fisica ``Galileo Galilei'', Universit\`a degli 
Studi di Padova, via Marzolo 8, 35131 Padova, Italy \break
$^4$International School for Advanced Studies, SISSA,
via Beirut 2-4, 34014 Trieste, Italy \break 
$^5$Nuclear and Astrophysics Laboratory, University of Oxford,
Keble Road, Oxford $\,$OX1 3RH, England  
}

\shortauthor{}
\shorttitle{Non-linear evolution of thermally unstable slim discs }


\acceptedline{Accepted . Received ;
  in original form }

\abstract 
{We are carrying out a programme of non-linear time-dependent numerical
calculations to study the evolution of the thermal instability driven by
radiation pressure in transonic accretion discs around black holes. In our
previous studies we first investigated the original version of the
slim-disc model with low viscosity ($\alpha = 0.001$) for a stellar-mass
($10M_{\odot}$) black hole, comparing the behaviour seen with results from
local stability analysis (which were broadly confirmed). In some of the
unstable models, we saw a violently evolving shock-like feature appearing
near to the sonic point. Next, we retained the original model
simplifications but considered a higher value of $\alpha$ ($=0.1$)  and
demonstrated the existence of limit-cycle behaviour under suitable
circumstances. The present paper describes more elaborate calculations
with a more physical viscosity prescription and including a vertically
integrated treatment of acceleration in the vertical direction.
Limit-cycle behaviour is still found for a model with $\alpha = 0.1$,
giving a strong motivation to look for its presence in observational data.
}

\keywords {accretion, accretion discs - instabilities}

\maketitle  

\section{Introduction}

The aim of the research programme, of which the present paper forms a
part, is to study the origin of the incipient instabilities operating in
accretion discs around black holes and to investigate their observational
consequences for diverse astronomical phenomena. For doing this, we use a
class of simple vertically-integrated, non-self-gravitating models of
transonic accretion discs. This is well justified because, despite the
ongoing development of increasingly sophisticated large-scale numerical
models, simple disc modelling still remains the central link between
theory and observations. Some of the results of disc phenomenology might
be truly fundamental, while others might have more limited domains of
applicability (Balbus \& Papaloizou, 1999). It would be particularly
valuable to determine the sensitivity of the disc solutions to the form of
the viscous stress and here we study how the global evolution of the
thermal instability driven by radiation pressure depends on the viscosity
prescription. This investigation follows on naturally from the overall
strategy of our research programme. The model treated in our previous
papers (Szuszkiewicz \& Miller 1997, 1998 - hereafter Papers I and II) did
not yet include several important non-local effects, and in particular, it
did not contain a diffusion-type formulation for the viscosity. Those
results represent a suitable standard reference for making comparison with
the results from our present calculations in which a diffusion-type
formulation for viscosity replaces the $\alpha p$ prescription used
previously.

The $\varphi r$ component of the viscous stress tensor, which is
responsible for the transport of angular momentum in a disc, is normally
written in the standard form
 $$ \tau_{\varphi r} = \rho \nu r {\partial\Omega \over \partial r}, 
\eqno (1) 
$$ 
 where $\rho$ is the density, $\nu$ is a kinematic viscosity coefficient
and $\Omega$ is the angular velocity of matter in the disc. However, the
relevant viscosity mechanism for accretion discs is clearly not molecular
viscosity (which is much too small to be of interest in these
circumstances) but instead is some sort of magnetic or turbulent
viscosity. For a general turbulent viscosity,
 $$ 
\nu_{_{turb}} \sim v_{_{turb}} \ell_{_{turb}},
\eqno (2)
$$ 
 where $v_{_{turb}}$ and $\ell_{_{turb}}$ are the characteristic
circulation velocity and length-scale for the turbulent cells (see Kato,
Fukue \& Mineshige 1998), and then it is common to write
 $$ 
\nu = \alpha H c_s, 
\eqno (3)  
$$ 
 where $c_s$ is the local sound speed, $H$ is the half-thickness of the
disc and
 $$ 
\alpha \sim \left( v_{_{turb}} \over c_s \right) \left( \ell_{_{turb}} 
\over H \right)
\eqno(4)
$$ 
 is a quantity which must be less than or comparable to unity if the
turbulence is subsonic with the turbulent cells being confined within the
disc. This treatment arises from discussions of viscosity associated with
hydrodynamic turbulence (von Weizs\"acker 1948) but a similar formula can
be appropriate for magnetic viscosity in some circumstances (Shakura \&
Sunyaev 1973; see also Hawley, Balbus \& Stone 2001).

The form of the viscosity plays a key role for the way in which the disc
responds to thermal fluctuations. An infinitesimal perturbation giving a
small rise in temperature would cause the disc to swell up and, according
to Eq.~(3), would have the tendency to make it more viscous giving
increased heating (although this also depends on how $\alpha$ changes in
response to the perturbation). At the same time, the cooling rate will
also become higher as a result of the higher temperature. If the increase
of cooling is greater than the increase in heating, the situation is
stable with the configuration returning back towards its initial state;
however, if the increase of heating is greater than the increase in
cooling, the situation is {\it thermally unstable} with the temperature
continuing to increase away from the initial value until the effect is
eventually saturated in a non-linear regime.  Whether this kind of thermal
instability can occur in practice for real discs depends on how fast the
viscosity increases as the temperature is increased and this is something
which is still under debate. Our view is that it is certainly of interest
to study the possible consequences of this thermal instability if it were
to arise and then to see whether evidence for it is revealed by
observations. This is the line being taken for the present programme of
work and, for simplicity, we follow the precedent of using the viscosity
formula (3) with constant $\alpha$ (which does allow for the instability
to occur if the accretion rate is high enough).

Within the framework of the above discussion, there is a further
simplification which is frequently made. If one considers a stationary
Newtonian Keplerian disc and makes a number of approximations (including
assuming vertical hydrostatic equilibrium and integrating analytically in
the vertical direction), it can be shown that expressions (1) and (3) 
lead to
 $$
\tau_{\varphi r} = - \alpha p ,
\eqno (5)
$$
 where $p$ is the pressure and $\alpha$ has been rescaled so as to remove
a numerical constant from the formula. This is the well-known $\alpha p$
prescription of Shakura \& Sunyaev (1973) which has been widely used as a
general approximate form for the viscous stress even under conditions
different from those for which the formula was derived.

Abramowicz et al. (1988), calculated a sequence of vertically-integrated
{\it slim} accretion disc models, using the $\alpha p$ viscosity
prescription but dropping the Keplerian flow condition and including the
effects of relativistic gravity in an approximate way by using the
pseudo-Newtonian potential of Paczy\'nski \& Wiita (1980). Doing this,
they found that if $\dot M$ (the accretion rate) is plotted against the
surface density $\Sigma$ for a fixed location in the disc, an S-shaped
curve is obtained. This suggested the possibility of having a limit-cycle
behaviour associated with a thermal instability driven by radiation
pressure. The results of non-linear time-dependent calculations performed
by Honma, Matsumoto \& Kato (1991) strongly indicated that such a
limit-cycle could indeed occur and the first complete solution of this
type was then subsequently obtained by Szuszkiewicz \& Miller (1998). This
finding, if confirmed by less approximate calculations, would have
important consequences for the interpretation of observations. A first
step towards checking its robustness is to replace the $\alpha p$
prescription by the one given directly by Eq. (1) together with expression
(3) for $\nu$. Doing this makes an important change in the mathematical
nature of the system of equations, introducing a parabolic diffusion-type
term into an otherwise hyperbolic system. 

Until now, little work has been directed towards studying transonic
accretion discs with this more general viscosity prescription because the
direct integration of the equations is extremely difficult (Hoshi and
Shibazaki 1977, Shibazaki 1978). The numerical results of Chen \& Taam,
(1993), who examined steady state accretion discs, demonstrated that the
S-shaped form of the $\dot M(\Sigma)$ curve is insensitive to the form
used for the viscous stress. Papaloizou \& Szuszkiewicz (1994b)
constructed models for the inner part of a transonic adiabatic accretion
disc, assuming constant specific angular momentum and including a full
treatment of the vertical structure, and for comparison purposes
constructed the corresponding one-dimensional viscous-disc models derived
under vertical-averaging assumptions (see also Papaloizou \& Szuszkiewicz,
1994a). They also investigated causality considerations connected with the
formulation of viscosity as a diffusion process and concluded that if the
specific angular momentum gradient is very small when the radial
velocities are significant, the corrections due to finite propagation
effects become small so that the Shakura \& Sunyaev (1973) treatment
should be reasonably good. Constraints arising from causality
considerations may be important though as far as the uniqueness of the
steady-state flow is concerned. Both $\alpha p$ and diffusive viscosity
prescriptions have been investigated by Artemova, Bisnovatyi-Kogan,
Igumenshchev and Novikov (2000). They found that the stationary solutions
for the two types of prescription are very similar for $\alpha \le 0.1$
but begin to differ at larger $\alpha$. They also showed that the
solutions with the diffusive shear stress have one singular point which is
always of the saddle type. A similar result had been obtained by
Papaloizou \& Szuszkiewicz (1994a) for transonic accretion discs with a
polytropic equation of state. The aim of our present study is to
investigate the effect of using a diffusive form of viscosity in global
non-linear time-dependent calculations of thermally unstable slim-disc
models.

The plan of the paper is as follows. In Section 2, we discuss the set of
basic equations used for our present calculations, highlighting the
changes with respect to the equations used in Papers I and II. The
modifications in the numerical treatment are described in Section 3. In
Section 4 we present the new results obtained with the revised computer
code and compare them with those obtained previously with the $\alpha p$
prescription. Section 5 contains comments and conclusions.

\section{What is new in the set of equations?}

The basic equations used for describing the evolution of the thermal
instability in an axisymmetric, non-self-gravitating, optically-thick disc
were presented in full detail in Paper I and the treatment which we will
be using for parts of the flow which are not optically thick was described
in Paper II. The assumptions and strategy used for the calculations have
been fully described in these earlier papers. Here we point out and
discuss two new features which we are now introducing into the set of
equations: the direct use of formula (1) for the viscous stress (giving
dependence on $\partial\Omega/\partial r$) and the introduction of a new
dynamical equation for the vertical acceleration (abandoning the previous
approximation of taking hydrostatic equilibrium to hold in the vertical
direction).

In cylindrical polar coordinates ($r$, $\varphi$, $z$) centred on the
black hole (having mass $M$), our basic hydrodynamical equations are as
follows. The conservation of mass equation
$$
{D \Sigma \over D t} = - {\Sigma \over r} {\partial \over \partial r}
\left( r v_r \right) 
\eqno (6)
$$
and the conservation of radial momentum
$$
{Dv_r \over Dt} =
- {1 \over \rho  }{\partial p \over \partial r} +
 {{\left( l^2 - l_{_K}^2 \right) } \over r^3} 
\eqno (7)
$$
are the same as in Papers I and II. We are using the following notation:
$D/Dt$ is the Lagrangian derivative given by
$$
{D \over Dt} ={\partial \over \partial t}
+ v_r {\partial \over \partial r }, 
\eqno (8)
$$
$\Sigma= \Sigma(r,t)$ is the surface density obtained by vertically
integrating the mass density $\rho$, $v_r = Dr/Dt $, $p$ is the total
pressure, $l=l(r,t)=rv_{\varphi}(r,t)$ is the specific angular momentum,
$l_{_K}$ is the value of $l$ for Keplerian motion in the pseudo-Newtonian
potential with $ v_{\varphi}=[ GMr/(r-r_{_G})^2 ]^{1/2}$, where $r_{_G}$
is the Schwarzschild radius of the black hole, and $\Omega_{_K} =
v_{\varphi}/r. $ Note that $v_r$ is negative for an inflow and that the
$\rho$ and $p$ appearing in Eq. (7) refer to values in the equatorial
plane.

The form of the azimuthal equation of motion used in the previous
calculations
$$
{Dl \over Dt}= -
{\alpha \over {r \Sigma}} {\partial \over \partial r} \left( r^2 pH
\right) 
\eqno (9)
$$
was obtained from
$$
{Dl \over Dt}= {1 \over {r \Sigma}} {\partial \over \partial r} 
\left( r^2 \ 2\int_0^H\tau_{\varphi r}dz\right) 
\eqno (10)
$$
using the $\alpha p$ viscosity prescription given by Eq. (5). Here, 
instead, we use the expression given directly by Eqs. (1) and (3):
$$
\tau_{\varphi r} = \rho \alpha c_s H r {\partial \Omega \over \partial r},      
\eqno (11)
$$
and Eq.~(10) then gives 
$$
{Dl \over Dt}=
{\alpha \over {r \Sigma}} {\partial \over \partial r}
\left(r^3 c_s H \Sigma {\partial{\Omega} \over \partial r}\right). 
\eqno (12)
$$
Note that, as mentioned earlier, the $\alpha$ appearing here is rescaled
with respect to that used previously with the $\alpha p$ prescription.
The value of the rescaling factor depends on the way in which the vertical
structure is treated. In Paper I, we described in detail how this was done
in our previous work and next we will give a modified discussion of how it
is done in the calculations of the present paper.

The vertical equation of motion can be written in the form 
$$ 
{D v_z \over Dt} = - {1 \over \rho }{\partial p \over \partial z} - 
	{\partial \Phi \over \partial z} + F_{zz}, 
\eqno (13)  
$$ 
where $\Phi$ is the pseudo-Newtonian potential given by $\Phi =
- GM/(R-r_{_G})$, with $R^2 = r^2 + z^2$, and $F_{zz}$ is a viscous force
which is set to zero throughout the present work. The strategy of the slim
disc approach is to proceed by deriving a vertically-integrated form of
this equation, making use of the fact that the disc is taken to be thin
enough so that the gravitational potential $\Phi$ can be expanded in the
vertical direction in powers of $z/r$ and only the lowest-order terms
retained. We then need two further assumptions about variations of
quantities in the vertical direction in order to be able to make the
integration. The pressure and density, $p$ and $\rho$, are taken to be
linked together (in the vertical direction) by a polytropic relation $p
\propto \rho ^{1+1/N}$, where $N$ is the polytropic index. Additionally,
we need to assume something about the vertical velocity and in the
previous work (in common with the preceding literature on slim discs) this
condition was provided by assuming vertical hydrostatic equilibrium to
hold so that $D v_z / Dt = 0$. The equation could then be vertically
integrated to give (after some manipulation and setting $N = 2$ following
Paczy\'nski \& Bisnovatyi--Kogan 1981):  
$$ 
\Omega_{_K}^2 H^2 =6 \, {p \over \rho}.  
\eqno (14)  
$$ 
This formula applies for strictly Newtonian gravity as well as with the
pseudo-Newtonian potential which we are using here. Inserting expression
(14) into the Newtonian argument leading to the $\alpha p$ formula, one
obtains that $\alpha_2$ (the $\alpha$ in the $\alpha p$ formula) is
related to the original $\alpha$, as in equation (1), by 
$$ 
\alpha_2 = {{3\sqrt{6}} \over {2}} \, \alpha_1 
\eqno (15) 
$$ 
We will use this relationship for calculating models within the new
approach which are ``equivalent'' to ones with a particular $\alpha$ in
the old formulation.

	In Paper I, we commented that neglecting the vertical acceleration
may sometimes be a rather poor approximation in some parts of the disc, as
suggested by two-dimensional studies (Papaloizou \& Szuszkiewicz 1994b),
and that investigating the effects of including vertical acceleration in a
consistent way was one of the important issues to be addressed 
subsequently. Contrary to our original strategy of introducing additional
modifications one by one, the acceleration term has been included here
together with the more general form for the viscosity. This was done
because we found that we could not obtain a numerically stable solution
with the revised viscosity law unless we also included the vertical
acceleration. We will discuss this further in Section 5.

	The simplest way of dealing with the acceleration term is to replace
the assumption of hydrostatic equilibrium with that of taking $z/H$ to be
constant along fluid flow lines. We note that while this is not in
accordance with the results of two-dimensional hydrodynamic simulations
(which reveal a complex circulating flow structure) it is consistent with
the general picture of the slim disc approach and therefore seems
appropriate here. Introducing this and then making the vertical integration
as before, one obtains
$$
{D V_z \over Dt} = 6{p \over \Sigma} - {GM \over (r-r_G)^2}
\left( {H \over r} \right),
\eqno (16)
$$
where $V_z$ is the vertical component of velocity for a fluid element at
the top surface of the disc.

	The introduction of the diffusive form of viscosity requires changes
also in the energy balance equation
$$
\rho T {DS \over Dt} = Q_{vis} + Q_{rad} . 
\eqno (17)
$$
Here $T$ is the temperature, $S$ is the entropy per unit mass, $Q_{vis}$ is
the rate at which heat is generated by viscous friction, now given by
$$
Q_{vis} = \rho \nu \left(r{\partial \Omega \over \partial r}\right)^2 , 
\eqno (18)
$$
and $Q_{rad}$ is the rate at which heat is lost or gained by means of
radiative energy transfer, here given by
$$
Q_{rad} = - {{F^-} \over {H}}, 
\eqno (19)
$$
with $F^-$ being the radiative flux per unit area away from the disc in the
vertical direction (we are neglecting heat conduction through the disc in
the radial direction). The vertically-integrated form of the energy balance
equation may be written in the following way
$$
{D T \over Dt} = {\alpha H r^2 T (\partial \Omega / \partial r)^2 \over
\sqrt{p/\rho}\ 0.67(12-10.5\beta)} - {T F^- \over 0.67pH(12-10.5\beta)}
$$
\vskip -0.4 truecm
$$
\hfill +{{(4-3\beta)} \over {(12-10.5\beta)}}{T \over \rho} {D\rho \over
Dt} \hskip 1.5 truecm \eqno (20)
$$
(Note that there is a typographical error in the equivalent equation of
Paper I -- there equation (17). On the right hand side there should be a
minus sign in front of the first term.)

If the medium is optically thick we use the expression
$$
F^- = {16 \sigma  T^4 \over  \kappa \rho H}, 
\eqno (21)
$$
where  $\sigma$ is the Stefan-Boltzmann constant and $\kappa$ is the
opacity. The opacity is approximated by the Kramers formula for 
chemical abundances corresponding to those of Population I stars
$$
\kappa = 0.34\, (1+ 6\times 10^{24}\rho T^{-3.5}) \ \ \
{\rm g}^{-1}\,{\rm cm}^2. 
\eqno (22)
$$
The thermodynamic quantities in the equatorial plane are taken to obey 
the equation of state
$$
p=k\rho T + p_r 
\eqno (23)
$$
where $p_r$ is the radiation pressure given by $p_r = {a \over 3}T^4$.

If the medium is not optically thick, we follow the approach of Hur\'e et
al. (1994) and write
$$ 
F^- = 6 { 4\sigma T^4 \over {{3\tau_{_R}  \over 2} + \sqrt{3} + 
{1\over \tau_{_P} }}}
\eqno (24)
$$
where $\tau_{_R}$ and $\tau_{_P}$ are the Rosseland and Planck mean
optical depths (equal to $\kappa_{_R} \rho H$ and $\kappa_{_P} \rho H$).  
The expression for the radiation pressure corresponding to this $F^-$ is
$$
p_r = {F^- \over 12c} \left( \tau_{_R}  + {2\over \sqrt{3}}\right) 
\eqno (25)
$$
For the Rosseland optical depth, we use the expression
$$
\tau_{_R} = 0.34 \, \Sigma \, (1+ 6\times 10^{24}\rho T^{-3.5}),  
\eqno (26)
$$
corresponding to the expression for $\kappa$ given by equation (22), while
for the Planck optical depth we use
$$
\tau_{_P} = {1\over 4\sigma T^4}(q^-_{br}) 
\eqno (27)
$$
where $q^-_{br}$ is the bremsstrahlung cooling rate given by
$$
q^-_{br} =1.24\times 10^{21} H\rho^2T^{1/2} \ \ \
{\rm erg}\,{\rm cm}^{-2}\,{\rm s}^{-1} 
\eqno (28)
$$
Further discussion of this has been given in Paper II.

Our aim here is to investigate the extent to which the inclusion of the
more physical viscosity prescription causes differences in non-stationary
behaviour with respect to that of the ``standard'' models calculated in
our previous papers.

\section{What is new in the numerical procedures?}

The equations presented in the previous section have been solved
numerically using a modified version of the Lagrangian hydrodynamics code
described in detail in Paper I, with the accreting matter being divided
into a succession of comoving radial zones and with the difference scheme
following the standard pattern for one-dimensional Lagrangian
hydrodynamics. In this section we discuss the modifications which have
been necessary in order to make a successful calculation for a model with
the new viscosity prescription and including the implementation of the
dynamical equation for motion in the vertical direction. 

The grid structure was modified from that used previously in order to give
better resolution where this was most needed. The gridding was extensively
tested to show that it was sufficiently fine to be well within the
convergent regime and, by experiment, we hope to have found a near optimal
compromise between accuracy and computational expense. The inner edge of
the grid was set at $r \approx 2.5 r_{_G}$ (whenever the innermost
comoving zone was entirely inside $2.5 r_{_G}$, it was discarded and the
grid was remapped following the procedure described in Paper I, using
piecewise cubic interpolation). The outer boundary was set at $r = 10^4
r_{_G}$, far enough away so that no perturbation from the inner regions
could reach as far as this during the time of the calculation. Conditions
there were essentially unchanging over this timescale, making it easy to
impose outer boundary conditions. For the model described in detail in
this paper, we used a grid of 400 comoving zones organised as follows: for
the first 59 zones, each one contained a mass 11 per cent larger than the
one interior to it; for the next 60 zones, the increase was 5 per cent;
for the rest of the disc, with exception of the last 60 zones, the
increase was 2.5 per cent and, finally, the last 60 zones had an increase
of 12 per cent. This was found to give satisfactory resolution and
accuracy.

The equations are solved for $\Sigma$, $v_r$, $l$, $V_z$ and $T$, as the
main dependent variables, with r and H being calculated from $Dr/Dt = v_r$ 
and $DH/Dt = V_z$ respectively. In our previous studies the time-step was
adjusted in accordance with the Courant condition and with two additional
constraints on the fractional variations of $\rho$ and $T$ in any single
time-step. Now that we have a parabolic term appearing in Eq.~(12), it is 
necessary to add a further time-step condition in order to maintain 
numerical stability:
$$
{\Delta t} \le { (\Delta r)^2 / (2 \nu ) }, 
\eqno (29)
$$
where $\Delta t$ and $\Delta r$ are the time-step and space-step
respectively. In the energy equation (20), which is solved implicitly for
$T$, the gradient of the angular velocity appearing in the viscous
heating term is no longer substituted by the Keplerian value as before.

When $r$, $\Sigma$, $T$ and $H$ are known, $\rho$ can be found directly
from $\rho = \Sigma/H$ while $p$ follows from the equation of state (23).
Previously, the solution for $p$, $\rho$ and $H$ was embedded in the
iteration loop for $T$. Introducing the dynamical equations for $V_z$ and 
$H$ significantly speeds up the code.

At every time-step, it is necessary to supply boundary conditions for
$v_r$ and $l$ at both the inner and outer edges of the grid. As before,
the boundary conditions at the inner edge are set by noting that $l$ is
essentially unchanging for infalling matter there, so that it can be taken
as constant in time for the innermost zone, and that the pressure gradient
term is having negligible effect there, so that it can be set to zero when
calculating $v_r$ at the inner edge. As noted above, the outer edge is 
located at such a large radius that conditions there are essentially 
unchanging over the timescale of the calculation and so $v_r$ is kept 
constant in time there while $l$ is calculated using the Keplerian 
condition at that point.

As initial conditions for the evolutionary calculations, we again used a
stationary transonic disc model calculated with the $\alpha p$ viscosity
prescription. The data from this were then transferred onto the
finite-difference grid used for the evolutionary calculations and the
$\alpha$ was suitably rescaled following Eq.~(15) for use in expression
(3). The quantities transferred were $r$, $v_r$, $\Sigma$, $T$ and $H$;
$l$ was then calculated on the grid, using the stationary version of
equation (7), and $V_z$ was calculated from the formula:
 $$ 
V_z = v_r {dH \over dr}.  
\eqno (30) 
$$ 

While treating the vertical equation of motion (13) by means of the
approximate form (16) is a great improvement over assuming hydrostatic
equilibrium in the vertical direction, it is still greatly simplified,
particularly in that it involves taking $Dv_z/Dt$ to depend linearly on
$z/H$ (which follows from $z/H$ being constant along flowlines). This
ceases to be reasonable when there are rapid dynamic changes in the
vertical direction such as those seen when the inner parts of the disc
deflate rapidly in the course of the evolution which we will be describing
in the next section. In reaching a new steady state after the deflation,
the vertical acceleration must inevitably deviate from $Dv_z/Dt \propto
z/H$. In order to keep the solution regular in these circumstances, we
introduced an artificial viscous term $Q_z \propto \rho V_z^2$ which
augments the pressure in the vertical direction when the vertical
compression is very rapid. This, to some extent, mimics the real behaviour
which would occur in these circumstances and is probably the best that can
be done within a vertically integrated approach although, of course, the
real answer is to go to a full 2D (or 3D) calculation! Related with this
problem is another one concerned with treating the stage in the immediate
aftermath of the deflation when a sharp density spike appears. The rapid
compression in the radial direction leads to a bunching of the Lagrangian
zones causing some of them to become so narrow that accuracy is lost. This
was combatted by adjusting the regridding routine to avoid having
ultra-narrow zones and simultaneously increasing the artificial diffusion
applied in the radial direction. The increased value of this artificial
diffusion coefficient is needed only for a very short time and, while its
effect is in the direction of causing a broadening of the spike, it is not
the main cause of the broadening seen which takes place over a much longer
time than that for which the diffusion is applied.

As noted in Paper I, the process of transferring the initial data onto the
finite-difference grid, together with the action of numerical noise which
inevitably enters during the calculation, is sufficient for introducing
suitable generalised perturbations in the model to trigger growth of
unstable modes which may be present.

\section{Limit cycle behaviour}

In Paper II we presented a complete limit-cycle solution for a model with
black hole mass $M=10M_{\odot}$, initial accretion rate $\dot M =0.06 \dot
M_{cr}$ and viscosity parameter $\alpha = 0.1$. ($\dot M_{cr}$ is the
critical accretion rate corresponding to the Eddington luminosity). Here
we investigate an equivalent model using the more physical viscosity
prescription given directly by equations (1) and (3) (for which
$\tau_{r\varphi} \propto \partial \Omega/\partial r$) and including the
effects of vertical acceleration by means of Eq. (16). The initial
equilibrium model is identical to that used in Paper II which, according
to the local stability criterion, is thermally unstable in a region
extending from $4.5\, r_{_G}$ to $17.5\, r_{_G}$. 

The main conclusion from our new calculations is that the overall behaviour
is very similar to that seen before (and described in Paper II). In
particular, we again see a succession of limit-cycles. We here present some
additional results and highlight some particular points of interest. The
figures presented here should be viewed in conjunction with those of Paper
II.

\beginfigure{1}
\vskip 15.3cm
\includegraphics{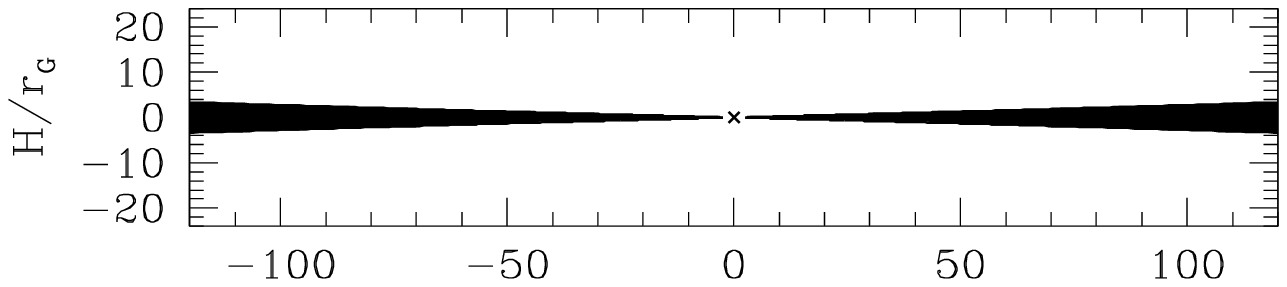}
\includegraphics{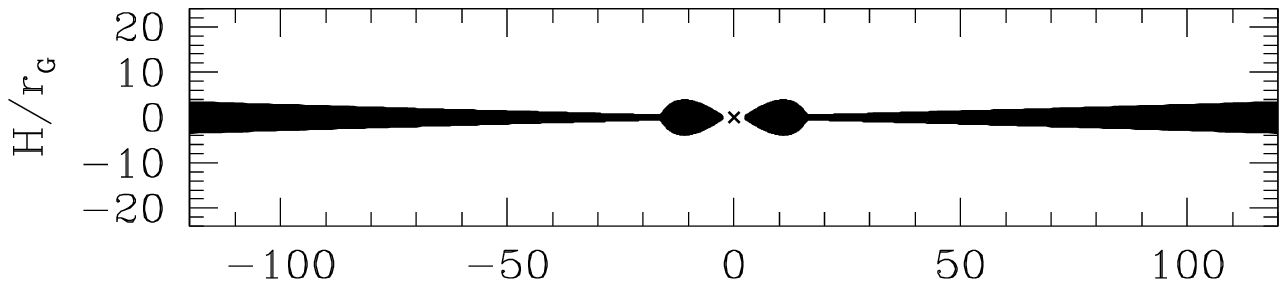}
\includegraphics{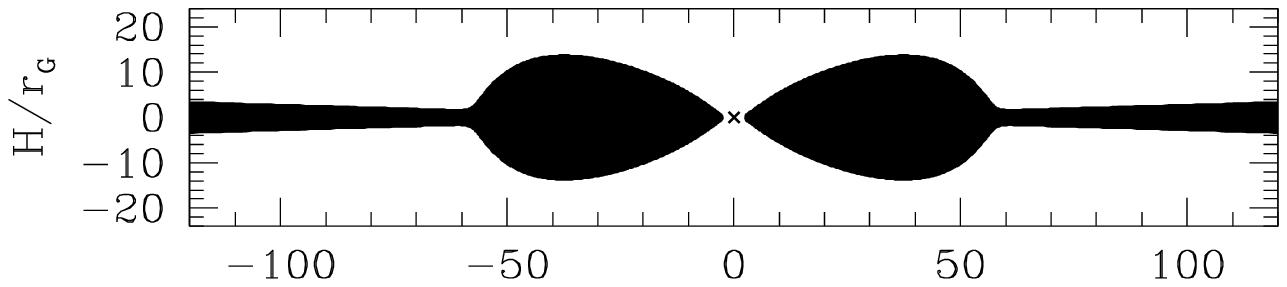}
\includegraphics{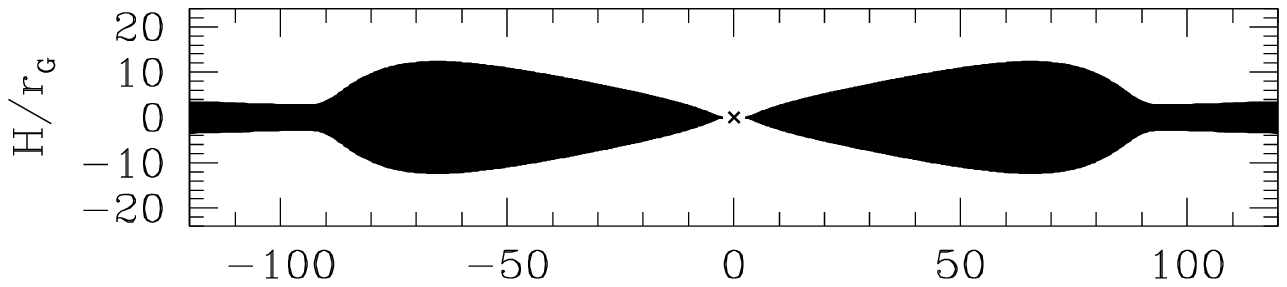}
\includegraphics{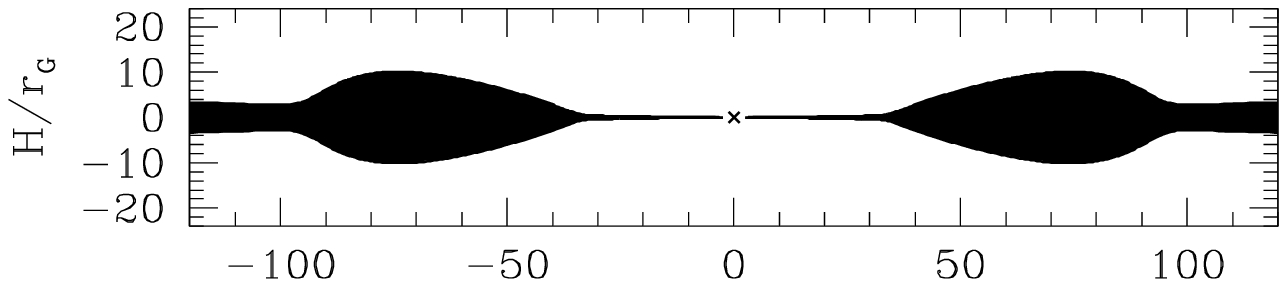}
\includegraphics{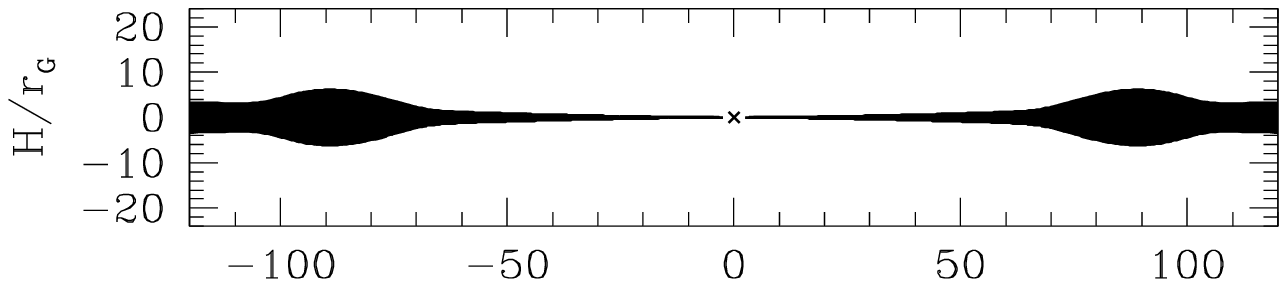}
\includegraphics{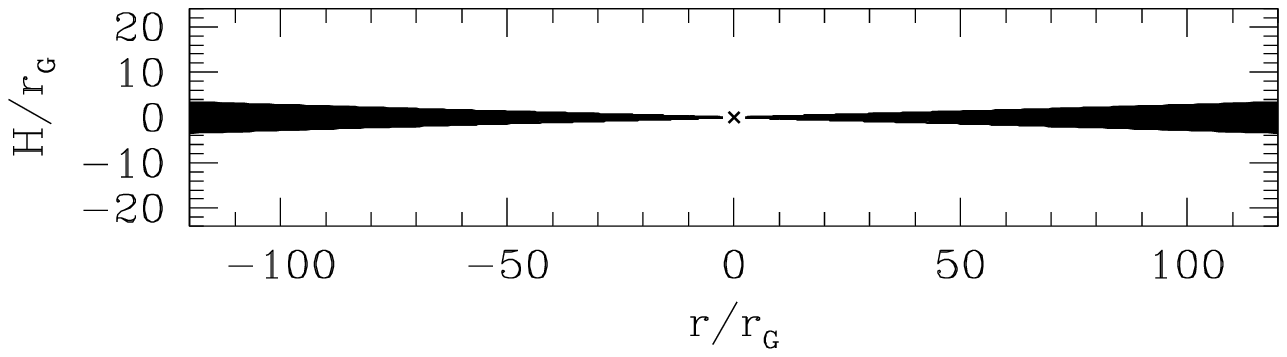}
\caption{{\bf Figure 1.}
The evolution of the geometrical profile of the disc during one full 
cycle. The sequence of panels shows the situation after 0, 2, 8, 16, 18, 
22 and $787\, {\rm s}$ respectively, starting from the beginning of the 
cycle. (Note that the outer boundary of the grid is at $r = 10^4\,r_{_G}$, 
far beyond the outer edge of the frame shown here.)
}
\endfigure

The evolution of the geometrical profile of the disc is shown in Fig.~1
and Figs.~2 -- 5 show the behaviour of the temperature $T$, the surface
density $\Sigma$, the effective optical depth $\tau_{_{eff}}$ and the
local accretion rate $\dot{m}$ at times corresponding to the first five
panels of Fig.~1. (No curves are drawn corresponding to the sixth panel in
order to avoid the figures becoming over-complicated.) In order to check
for possible long-term variations, we continued the calculation for a
total of twenty complete cycles; the results shown in Fig.~1 are for a
representative cycle -- the 17th.

\beginfigure{2}
\vskip 8.0cm
\includegraphics{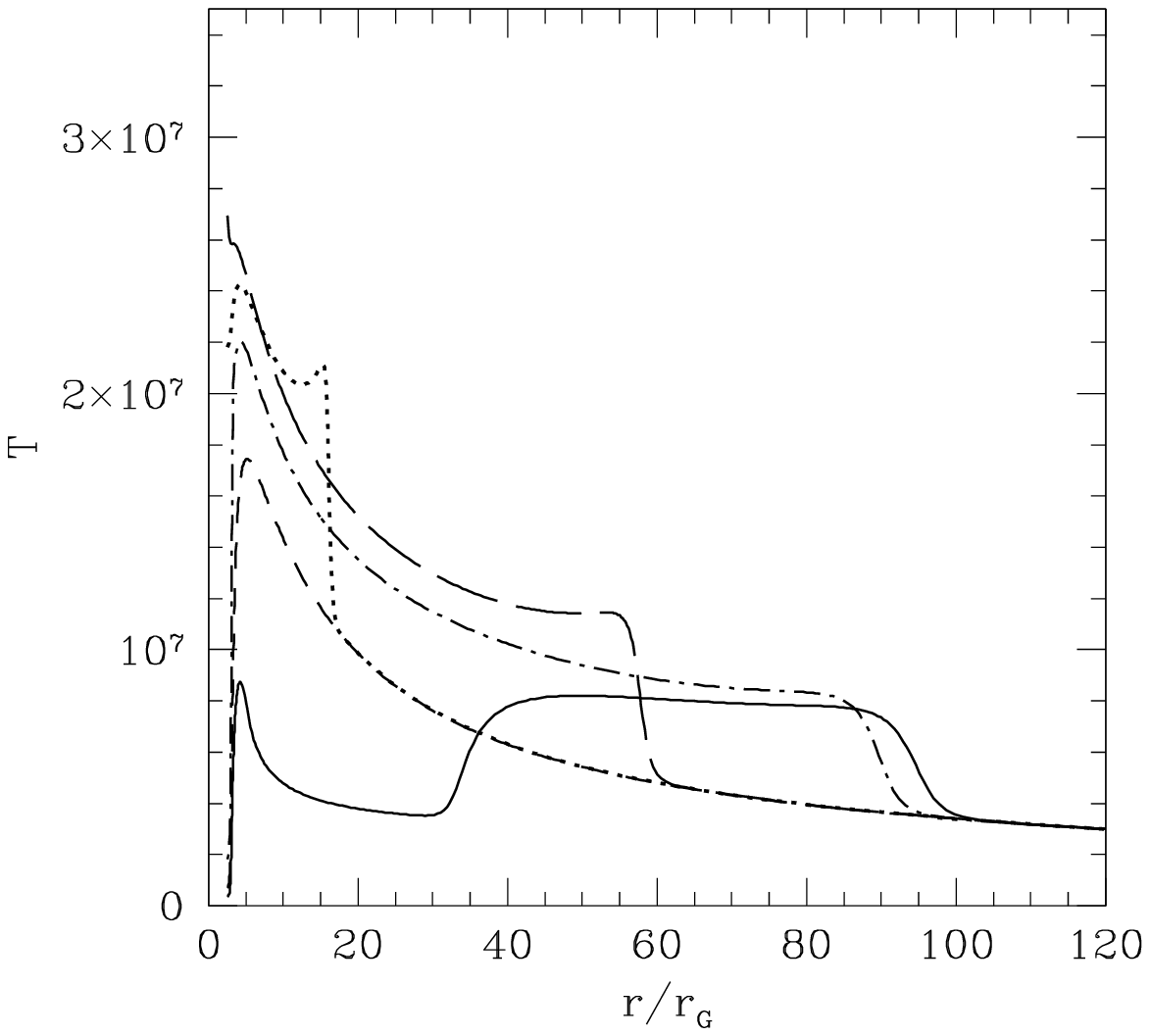}
\caption{{\bf Figure 2.} The temperature $T$, measured in degrees kelvin,
is plotted against $r/r_{_G}$ at successive times during one full cycle.
The short-dashed curve shows the temperature at the beginning of the cycle
($t=0 \, {\rm s}$), the dotted curve is for $t=2 \, {\rm }$s, the dashed
curve is for $t=8 \, {\rm s}$, the dot-dashed curve is for $t=16 \, {\rm
s}$ and the solid curve is for $t=18 \, {\rm s}$.
}
\endfigure  

\beginfigure{3}
\vskip 8.5cm
\includegraphics{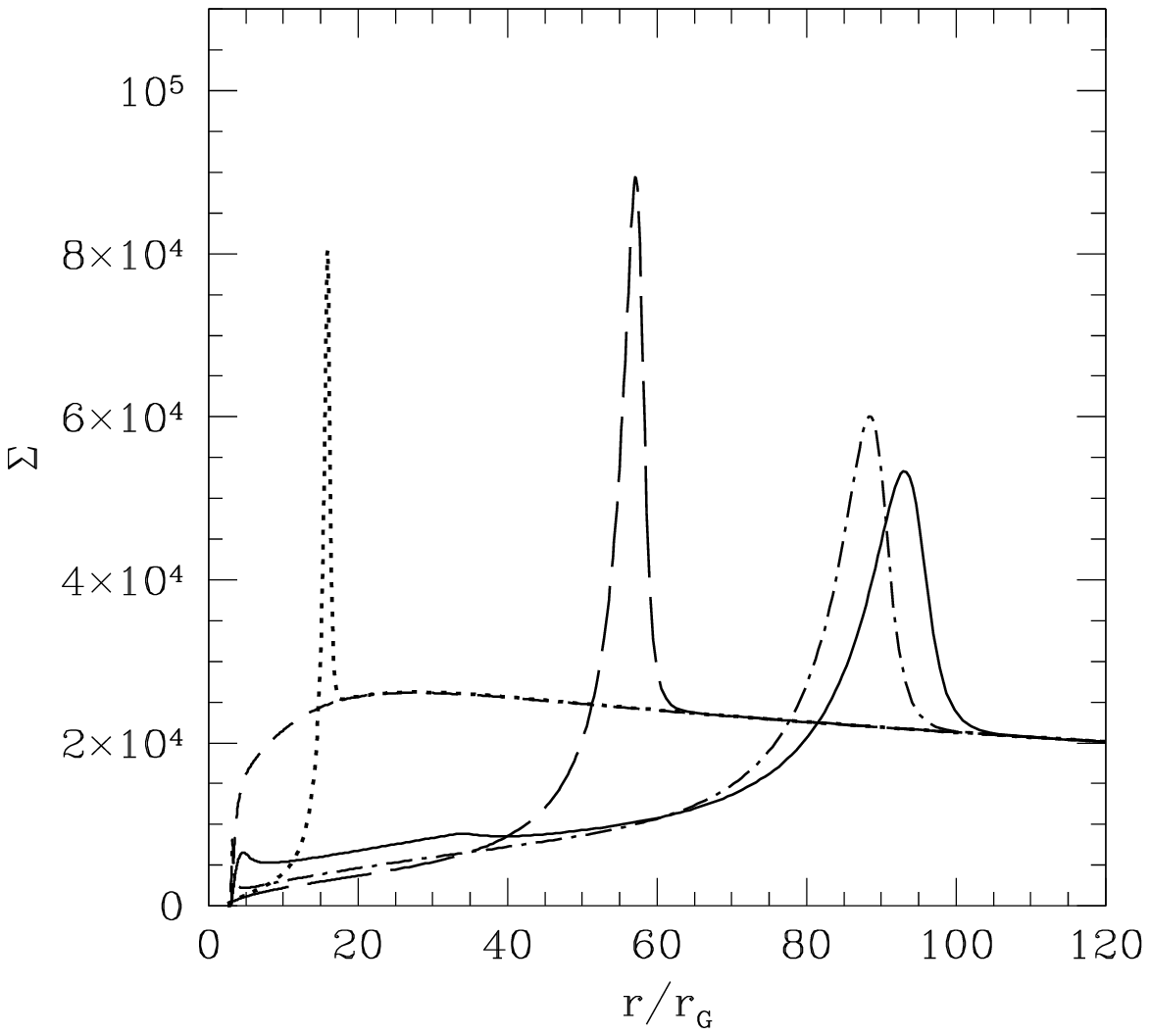}
\caption{{\bf Figure 3.} The surface density $\Sigma$, measured in units 
of g$\,$cm$^{-2}$, is plotted against $r/r_{_G}$ at the same times as 
shown in Fig.~2.}
\endfigure  

\beginfigure{4}
\vskip 8.0cm
\includegraphics{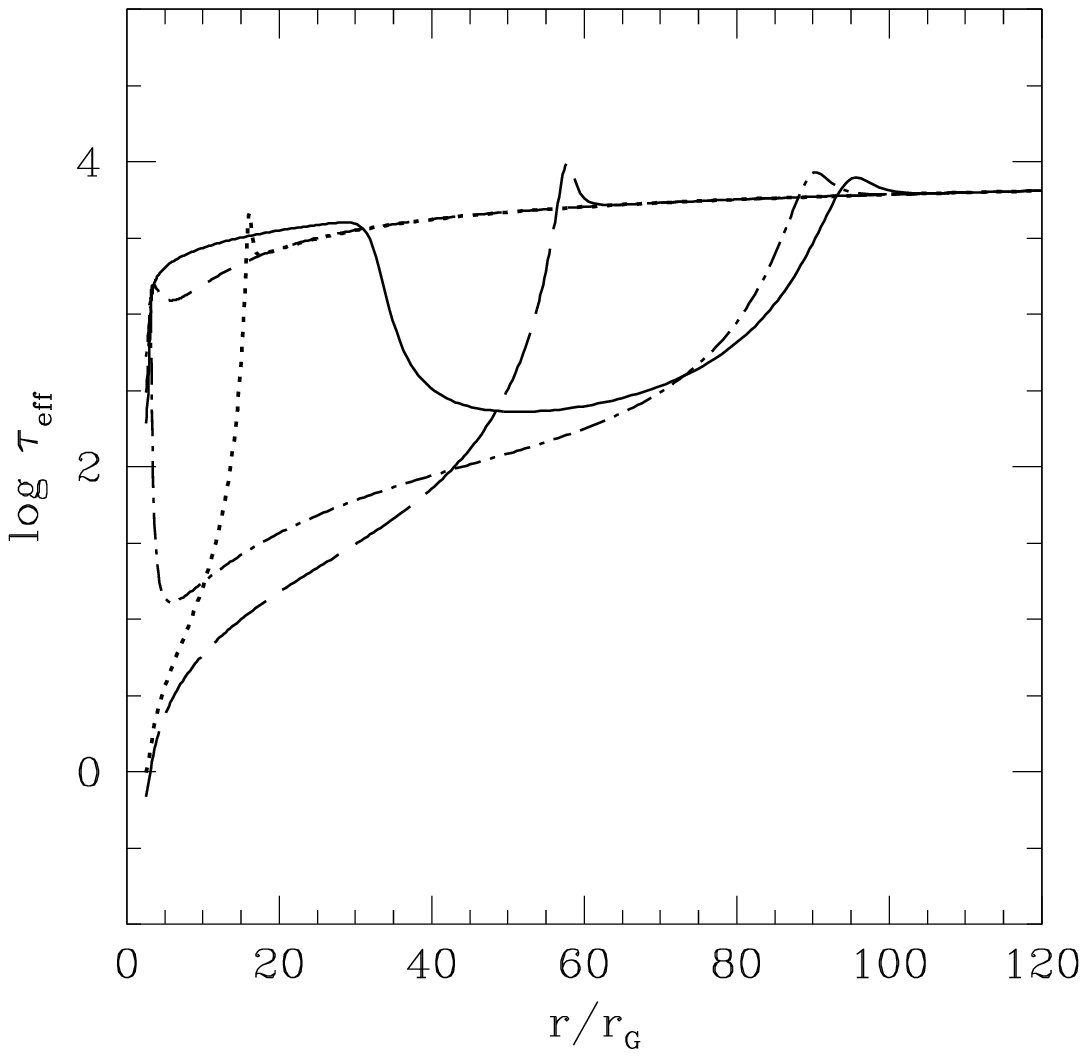}
\caption{{\bf Figure 4.} The effective optical depth $\tau_{_{eff}}$ is 
plotted against $r/r_{_G}$ at the same times as shown in Fig.~2.
}

\endfigure  

\beginfigure{5}
\vskip 8.5cm
\includegraphics{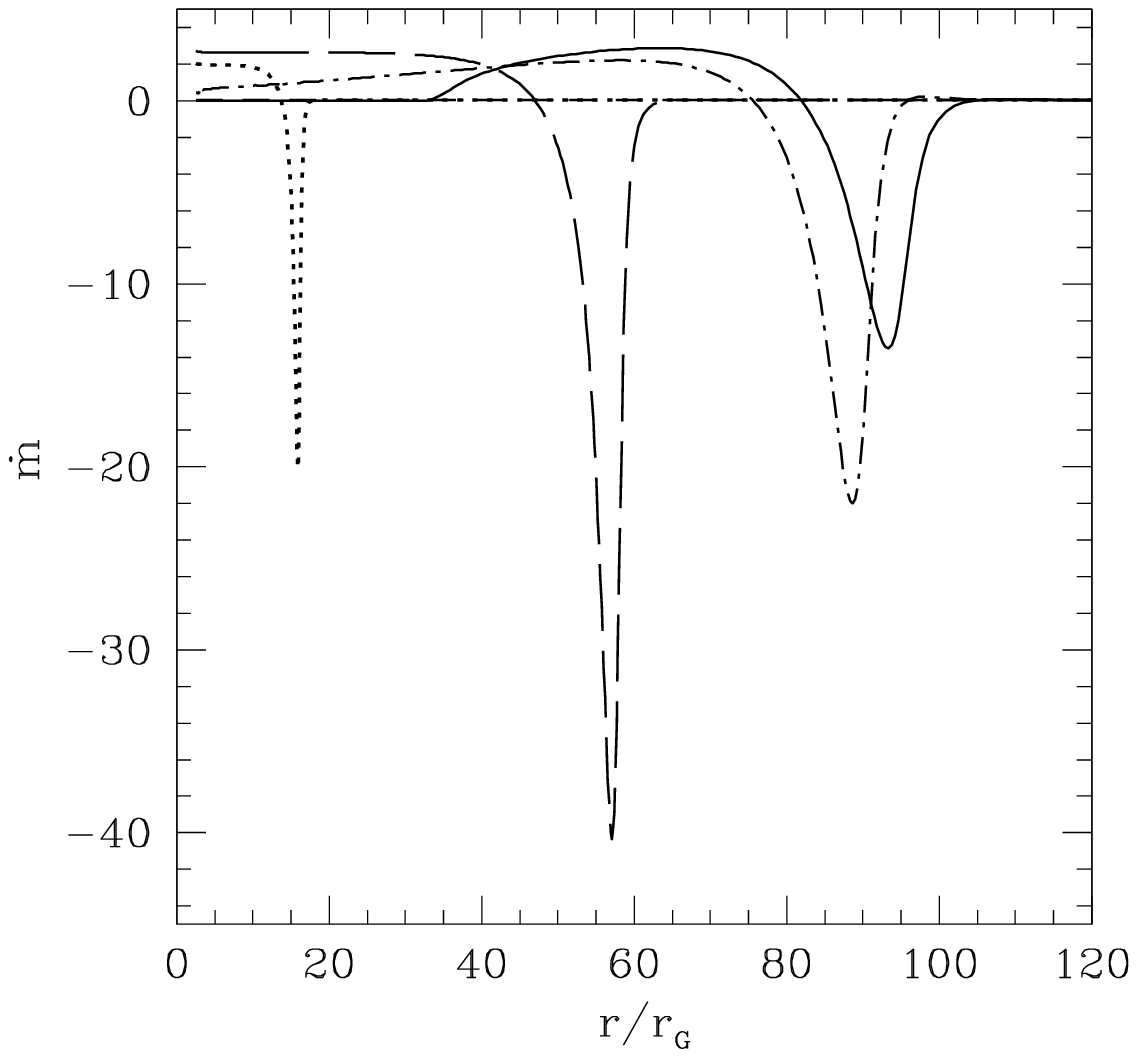}
\caption{{\bf Figure 5.} The local accretion rate $\dot m(r)$, measured in
units of the critical accretion rate $\dot M_{cr}$, is plotted against
$r/r_{_G}$ at the same times as shown in Fig.~2. Negative values of $\dot
m$ signify an outflow, $(\dot m = -2\pi r \Sigma v_r)$.}
\endfigure

The first panel of Fig.~1 shows the disc just before the start of the
cycle (which we define to be the moment when the instability first begins
to exponentiate). As the instability sets in, the temperature in the
unstable region rises rapidly above its stationary value, increasing the
contribution of the radiation pressure by nearly an order of magnitude.
The heated material expands in all directions, pushing inner material into
the black hole, expanding the disc vertically and launching a compression
wave out through the disc which sweeps matter ahead of it leaving a
semi-evacuated region behind. This is the stage reached by 2 seconds into
the cycle, corresponding to the second panel in Fig.~1. At the start of
the cycle, the effective optical depth $\tau_{_{eff}}$ is greater than 100
everywhere but it becomes smaller in the inner parts of the disc as the
material there expands, falling below 10 for $r < 9\, r_{_G}$ by the
present stage. The behaviour of this and the other quantities is shown by
the corresponding curves (marked with dots) in Figs.~2 -- 5. (Note that
despite their sharp appearance, all of these curves are, in fact,
perfectly continuous and well-resolved on the grid as can be seen if they
are viewed with a larger spatial resolution.) As the outgoing wave
progresses further, the temperature peak is progressively reduced in
magnitude although it remains at a level well above that of the initial
model. The outgoing wave is heating the material through which it passes,
causing the part of the disc internal to it to swell up further as shown
in the third panel of Fig.~1. The region with very low effective optical
depth ($\tau_{_{eff}} < 10$) is now extending out to $15\, r_{_G}$. The
compression wave is pushing material strongly outwards away from the black
hole (note, in Fig.~5, the large negative $\dot m$ associated with it)
while, behind it, matter is falling inwards at a rate which is far above
that in the stationary state ($\dot m = 0.06$) and is even well above the
critical accretion rate $\dot m = 1$. In this initial phase of the
evolution, the propagating compression front divides the disc into two
parts. The inner part has a toroidal structure and is characterised by
high temperature, low density, $\tau_{_{eff}} < 100$, and a
nearly-constant supercritical accretion rate, while the outer part is
basically similar to the original unperturbed stationary disc (although,
as we will see later, it is actually far from being a stationary flow for
cycles after the first one). The width of the transition zone between
these two parts increases as the front moves outwards and has reached $20
\, r_{_G}$ by 8 seconds into the cycle, corresponding to the third panel
of Fig.~1.

In order for the outgoing wave to continue its propagation, it is
necessary that the material behind it should remain in a hot state. Once
the wavefront has moved beyond the linearly-unstable region, it becomes
progressively harder for the temperature to be maintained in the high
state and eventually the front starts to weaken, the temperature drops in
the innermost part of the disc and the material there deflates, collapsing
down into the equatorial plane. Panel 4 of Fig.~1 shows the situation at
the moment when this happens, $16 \, {\rm s}$ into the cycle, and in the
subsequent panels 5 and 6 one can see the deflation spreading out through
the disc. During the initial collapse, the volume density $\rho$ rises
dramatically at the place where the collapse occurs, growing by a factor
of $10^3$ in less than a second and forming a sharp spike which oscillates
and then starts to broaden progressively. (This is a feature which is easy
to follow in animations of the data but is hard to show clearly with a
static graph.) At the time of panel 4, the accretion rate is decreasing
towards the black hole and the optical depth has increased considerably in
the inner parts of the disc.

Following this rapid change, the temperature continues to decrease and
drops to a low state well below that of the initial model. By the time
corresponding to panel 5 of Fig.~1 ($18 \, {\rm s}$ after the beginning of
the cycle), the compression front is seriously weakening and a growing
region of the inner part of the disc has deflated. As can be seen from
Fig.~3, there is very little matter inward of the compression front at
this stage but the process of refilling the inner part of the disc is now
beginning, as the accretion flow near to the black hole is almost turned
off. The disc now consists of three different parts. The innermost part
forms a geometrically thin disc-like configuration with the temperature
and surface density being lower than those of the initial model and with a
very low accretion rate. The size of this part is progressively
increasing; at $18 \, {\rm s}$ it extends out to just beyond $30 \,
r_{_G}$. The middle part is what remains of the toroidal structure. This
is still at a higher temperature than in the initial model, is
under-dense, has a lowered optical depth and has a high inward accretion
rate. The outermost part is still in a similar state to that of the
initial model but has been compressed somewhat by the still outward-moving
compression front.

At the time of frame 6 of Fig.~1 ($22 \, {\rm s}$ after the beginning of
the cycle), the first section of the disc (geometrically thin and
under-dense) extends out to $65 \, r_{_G}$, the compression front has
almost died and the remaining toroidal structure is shrinking away.

This is the end of the ``outburst'' part of the cycle (which has proceeded
on the thermal timescale). What happens next is a slow and uneventful
refilling (and reheating) of the inner part of the disc which occurs on
the much longer viscous timescale. Finally ($787 \, {\rm s}$ after the
beginning of the cycle -- Fig.~1, 7th panel) the disc has returned to
essentially the same state as it was in at the beginning of the cycle. The
thermal instability then appears again and a new cycle is initiated.

\beginfigure{6}
\vskip 8cm
\includegraphics{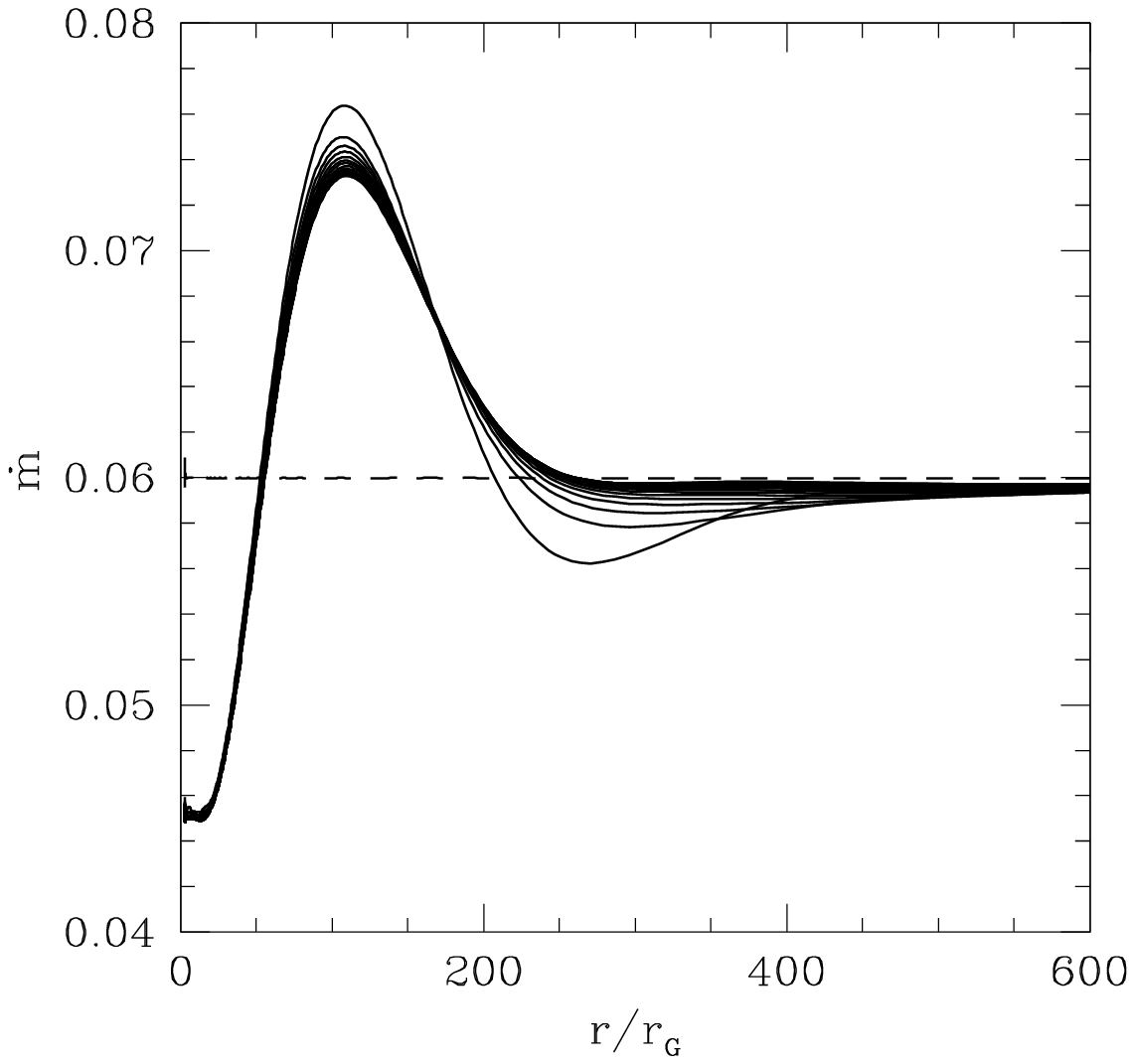}
\caption{{\bf Figure 6.}
The local accretion rate profile at the beginning of successive cycles 
(note that the vertical scale is very different from that of Fig.~5). 
The convergence process is rather slow; the final converged profile shows 
$\dot{m}$ varying significantly with $r$ and so represents a genuinely 
non-stationary solution.
}
\endfigure

We have calculated the time evolution of the model up to the beginning of
the twenty-first cycle. The cycle period ($\sim 787 \, {\rm s}$) is
slightly longer than the $780 \, {\rm s}$ found in our previous study, and
it was seen to be gradually increasing with time (albeit with extremely
small fractional changes) until a fully relaxed solution was finally
obtained. In connection with this, it is important to stress that after
the initial time, at which data for a stationary solution was specified,
the configuration never again passed through a stationary state and was
not even close to doing so. Figure 6 shows the profile of $\dot m$ at the
beginning of successive cycles (just before the instability starts) with
the initial model being shown with the dashed line and the subsequent
curves being in order of progressively decreasing amplitude. Note that for
the fully relaxed case, $\dot m$ has a minimum of 0.045 (near to the black
hole) and a maximum of 0.073 (just outside $100\,r_{_G}$) and so is far
from being constant at 0.06 as was the case for the initial model. In view
of this, it is clear that a number of cycles would be needed before a
relaxed solution could be obtained and this is the main reason for the
progressive change seen in these $\dot m$ curves although there is also
some evidence of smoothing related to the parabolic prescription for the
viscosity.

\beginfigure{7}
\vskip 8cm
\includegraphics{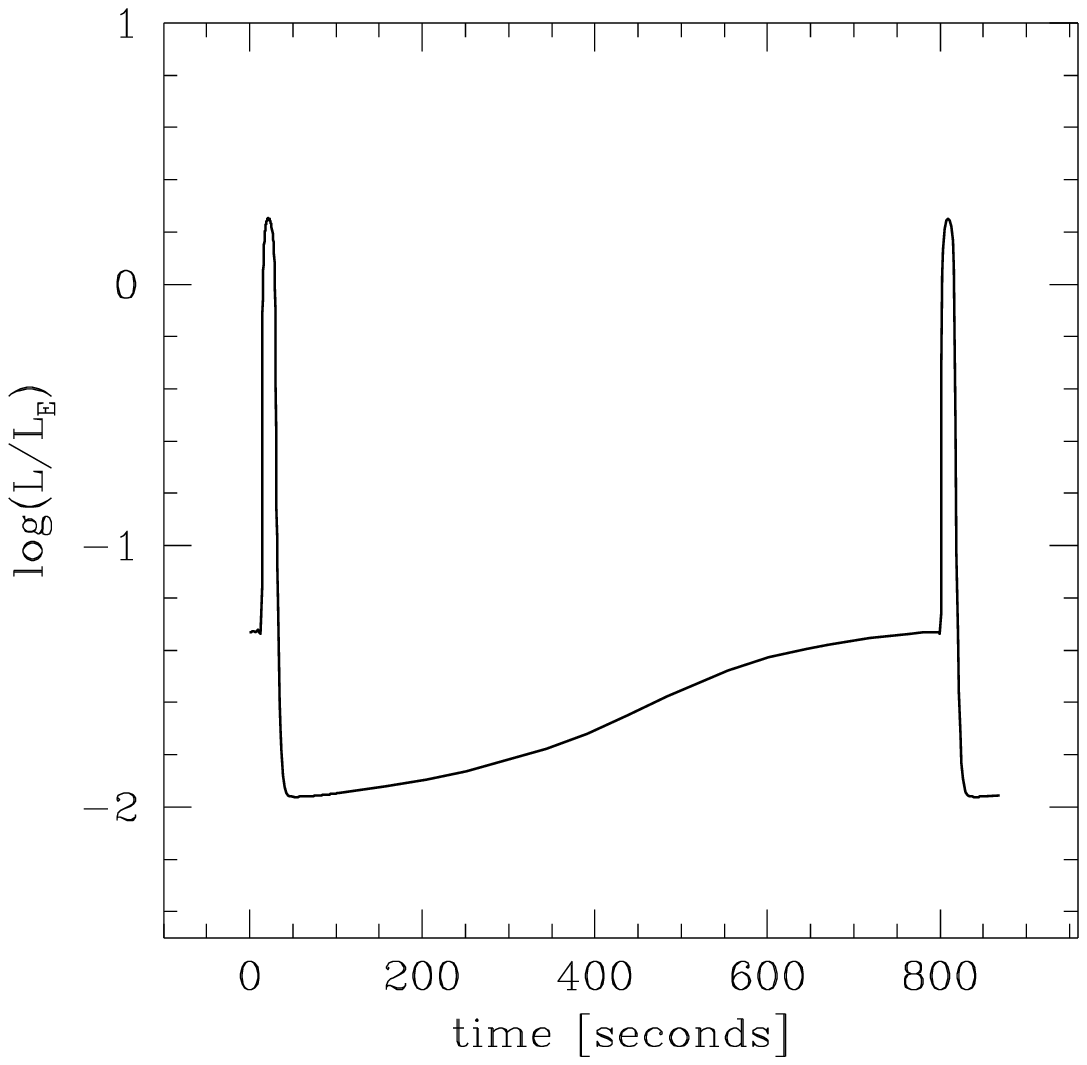}
\caption{{\bf Figure 7.}
Variation of the bolometric luminosity of the disc expressed in units of the
Eddington luminosity.
}
\endfigure

Integrating the radiated flux per unit area over the disc at successive
times, we have obtained the bolometric light curve shown in Figure 7. The
disc luminosity exhibits a burst-like time variation with a burst duration
of about $20 \, {\rm s}$ and a quiescent phase lasting for the remaining
$767 \, {\rm s}$ of the cycle. The amplitude of the variation is around
two orders of magnitude: at the maximum, the luminosity is approximately
40 times larger than it was immediately before the outburst and after the
peak it then drops below the pre-burst value by a factor of about 4.5.  
During the following quiescent phase, it then gradually increases again
until the onset of the next outburst.

The bolometric luminosity is not a directly observed quantity, and so even
if we expect that most of the energy in galactic black hole candidates will
be radiated in the X-ray band, it is necessary to calculate the spectrum
emitted from the disc and the light curves in the particular wavebands, in
order to perform any detailed comparison with observations. This will be
discussed in subsequent papers.

\section{Discussion and conclusions}

In the previous calculations presented in Papers I and II, which we use
here for comparison with the results of the present more elaborate study,
we investigated the simple slim-disc model with all of its associated
assumptions and approximations. Two different values of the viscosity
parameter $\alpha$ (low - 0.001 and high - 0.1) were considered there. In
the present work, two important new aspects have been introduced: the
diffusion-type formulation for the viscosity and the dynamical treatment
of motion in the vertical direction. We now examine some features
discussed before for the previous formulation to see how these are changed
with the new one.

\beginfigure{8}
\vskip 8cm
\includegraphics{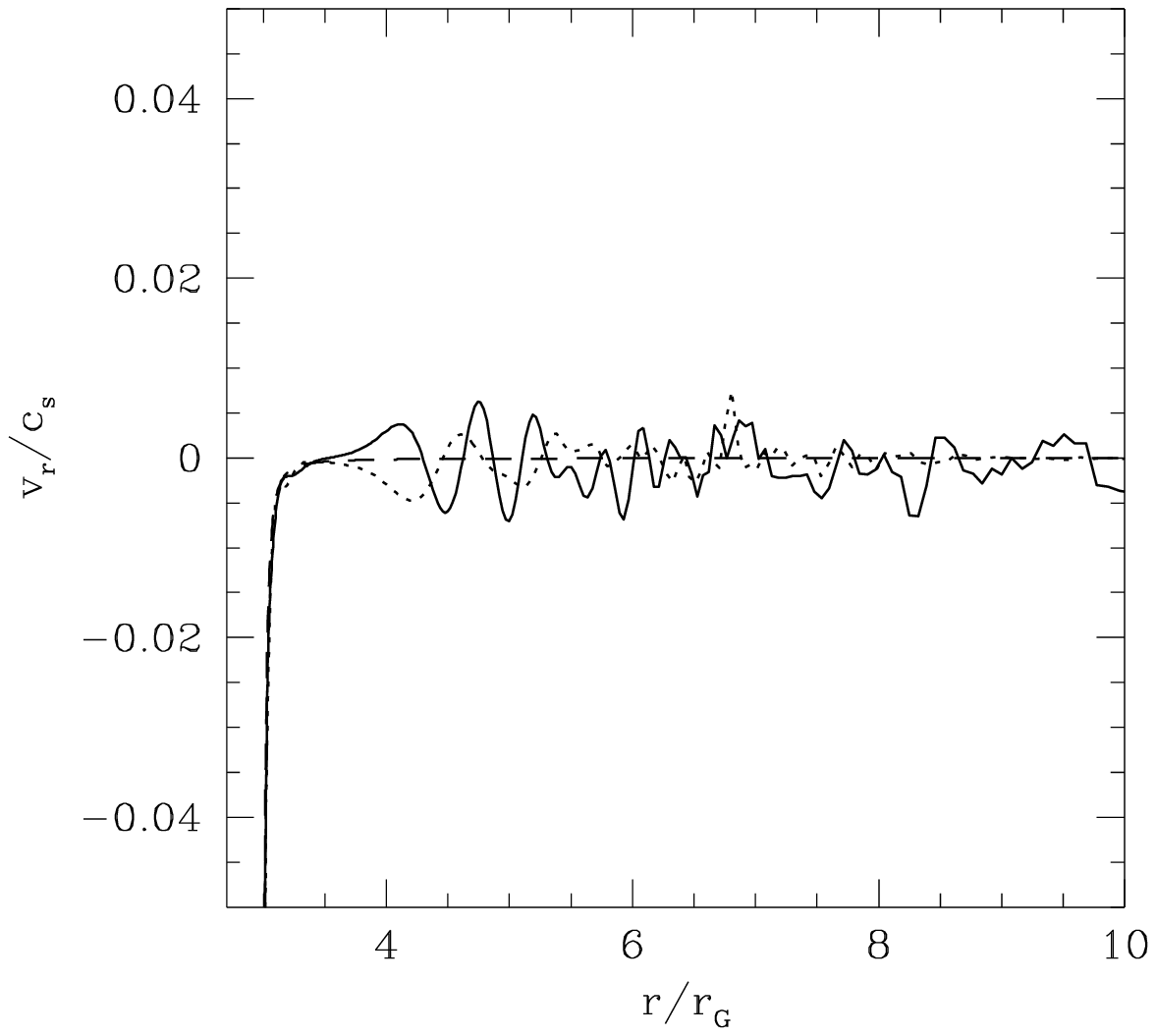}
\caption{{\bf Figure 8.}
Numerical noise which appears when the artificial diffusion term is
removed. These velocity curves are for the model with $\alpha =0.001$ and
$\dot m =0.01$ at three different times: $t=0 \, {\rm s}$ (dashed line),
$t=76 \, {\rm s}$ (dotted line) and $t=1881 \, {\rm s}$ (solid line). Note
that the vertical scale is very much expanded with respect to Fig.~11 of
Paper I so that the oscillations can be seen.
}
\endfigure

In Paper I we observed that the supposedly stable model with $\alpha
=0.001$ and $L=0.01\,L_{_E}$ did not remain unchanging but instead
developed an instability in the region $3 - 6\,r_{_G}$ after $10-20\, {\rm
s}$ of the evolution (see Figure 11 in Paper I). This appeared first at
around $4.7 - 4.8\,r_{_G}$ and then grew rapidly, causing termination of
the calculation. The high-$\alpha$ model of Paper II was subject to the
same kind of problem. We found that the instability could be suppressed by
adding a low level of numerical diffusion to the radial equation of motion
and this suggested that additional physical diffusive terms, which were
not included in the simple version of the model, might play a key role in
avoiding this type of instability. Following our strategy, we repeated the
calculation for the thermally-stable model studied in Paper I, using the
new version of the code and leaving out the numerical diffusion term. We
found that the oscillations do appear again (see Figure 8) but this time
they remain bounded at a low level and do not cause termination of the
evolution. In particular, we note that they are not trapped around their
point of origin but are able to propagate freely outwards. In the case of
the high-$\alpha$ model studied here, these oscillations interfere with
the developing thermal instability which is, of course, undesirable and
so having studied their behaviour under these new circumstances, we then
re-introduced the low level of artificial diffusion as before in order to
suppress them.

Next we consider the violently unstable feature seen in the calculations
presented in Paper I for models with low viscosity and luminosities
between $0.09\,L_{_E}$ and $1\,L_{_E}$. This instability manifested itself
as a dramatic shock-like feature which appeared near to the sonic point
and then grew rapidly, destroying the calculation. Similar behaviour was
seen in equivalent calculations made by Ryoji Matsumoto (private
communication)  using a different numerical method. The nature of this
behaviour is still under study. If not avoided or saturated, it would lead
to disruption of the disc meaning that models like these would not be
suitable for describing persistent astronomical flow configurations. We
made a detailed investigation of the effect of introducing different forms
of artificial viscosity or diffusion and found that none of these could
act to limit the instability unless they were increased to levels which
did not seem reasonable. We emphasise that our numerical code has no
difficulty in dealing with ordinary shock structures and, indeed, in Paper
II we showed a disc evolution with strong shocks present in the supersonic
part of the flow. In fact, despite first impressions, the dynamically
growing feature turns out not actually to be a shock but, instead, is a
transient feature across which the Rankine-Hugoniot conditions are not
satisfied. It is therefore not particularly surprising that methods
designed for handling shocks were not successful in dealing with this.
By adding a large enough amount of viscous dissipation, the disruptive
nature of the feature can indeed be removed but our feeling was that
introducing so much diffusion without physical motivation was unwarranted
(as we said in Section 4 of Paper II). At that stage, we wanted to wait
for the introduction of a $(\partial \Omega/ \partial r)$ viscosity to see
the effect of that before proceeding further. Perhaps a good description
of the feature is that it is like a wave washing up against a sea-wall
with the density gradient near to the sonic point playing the role of the
wall. When sufficiently high artificial diffusion was added, the wave was
seen to fall back from the wall and then to wash up against it again
repeatedly but without disrupting the rest of the solution. The
introduction of the additional features in our current code has not turned
out to be sufficient, in itself, to avoid the instability destroying the
solution unless similarly high levels of artificial diffusion continue to
be added.

Our original plan was to perform a systematic study of disc evolution with
additional effects being added one at a time but here we have gone against
this by introducing, at the same time, the diffusive ($\partial \Omega /
\partial r$) viscosity prescription and the dynamical treatment of motion
in the vertical direction. The reason for this was because of a
computational difficulty which we encountered with the new viscosity
prescription at the stage when the vertically-expanded part of the disc
started to deflate. All of our attempts to calculate through this stage
with the $\partial \Omega / \partial r$ viscosity were unsuccessful until
we decided to introduce also the vertical acceleration, at which point the
problem disappeared. It seems that the approximate nature of the $\alpha
p$ prescription somehow compensated the similarly doubtful use of the
condition of vertical hydrostatic equilibrium.

In our calculations, we have reported two main types of time-varying
behaviour for models undergoing the thermal instability: for our higher
value of $\alpha$ ($= 0.1$), we have seen limit-cycles, while for low
$\alpha$ ($= 0.001$), we saw a seemingly catastrophic instability.  
Takeuchi \& Mineshige (1998) have presented extremely interesting results
from an $\alpha p$ calculation for very high $\alpha$ ($=1$) in which they
see an evolution which starts in the same way as our model which makes
limit cycles but, after the general deflation, a small inflated, optically
thin region remains at the inner edge of the disc (inward of $r = 5 \,
r_{_G}$) which seems to be a persistent ADAF-like feature. This blocks the
possibility of starting a new outburst cycle. The authors give a
persuasive argument for why this should occur, indicating that there
should be a difference in behaviour for models with $\alpha > 0.3$, for
which the expanded states are fully optically thin (our model with $\alpha
= 0.1$ has considerably reduced optical depth in the expanded region but
only has $\tau_{_{eff}} < 1$ briefly and then only for a very small region
at the inner edge of the disc). When we have, in the past, tried to make
calculations for $\alpha = 1$, we experienced difficulty in keeping the
code numerically stable (and there is, perhaps, evidence of some similar
problem in the graphs shown by Takeuchi \& Mineshige, although we stress
that we find their results believable, particularly in view of the
supporting arguments which they present). Clearly, it is now of interest
for us to return to those calculations and see whether we can confirm the
results of Takeuchi \& Mineshige. In particular, it will be interesting to
see whether the persistent ADAF feature would survive changing the
viscosity prescription to the diffusive form.

In conclusion: the main result of our present paper concerns the effect of
replacing the $\alpha p$ viscosity prescription by the more physical
diffusive form when calculating the global evolution of the thermal
instability, driven by radiation pressure, for a vertically-integrated
model of a non-self-gravitating transonic accretion disc in the
high-$\alpha$ case ($\alpha = 0.1$). We find that the evolution remains
cyclic in character and is very little changed from that seen before. This
provides some strengthened motivation for saying that the type of disc
behaviour seen in our calculations might truly have some fundamental
importance. It is then natural to proceed with studying the possibility
that its effects might be present among the range of different variability
patterns observed for accreting compact objects. The transonic discs
described here with accretion rates larger than a certain value (which
depends on the mass of the central black hole, $M$, and the viscosity
parameter, $\alpha$) are not stationary but show a cyclic behaviour which
can be characterised by the period of the cycles, the durations of the
bursting phase and of the quiescence, and the amplitude and shape of the
burst. For different input parameters: $M$, $\alpha$, $\dot m$, we should
expect to see different characteristics. A straightforward comparison with
XTE observations, for example, can lead to identification of sources where
the thermal limit-cycle behaviour may be responsible for observed
variabilities. So far, there is only one source for which this mechanism
has been proposed: GRS 1915+105 (Belloni et al. 1997) and the situation
for that is still under investigation. Good quality data are now available
for making a detailed confrontation with clear model predictions.

\section*{Acknowledgments}

We gratefully acknowledge financial support from the Polish State
Committee for Scientific Research (grant KBN 2P03D01817), the Italian
Ministero dell'Uni\-ve\-rsi\-t\`a e della Ricerca Scientifica e
Tecnologica, the Italian INFN and the European Union (under the EU
programme ``Improving the Human Research Potential and the Socio-Economic
Knowledge Base'': RTN contract HPRN-CT-2000-00137).

\bye